\documentclass[%
 reprint,
 amsmath,amssymb,aps,
longbibliography]{revtex4-2}

\usepackage{graphicx}
\usepackage{dcolumn}
\usepackage{bm}

\usepackage{enumitem}
\usepackage{soul}
\usepackage{ulem}
\usepackage{xcolor}
\usepackage{multirow}
\usepackage{url}

\usepackage{amsmath}
\usepackage{amsthm}
\usepackage{amssymb}
\usepackage{amsfonts}
\usepackage{enumerate}
\usepackage{braket}
\usepackage[utf8]{inputenc}
\usepackage[top=1 in,bottom=1in, left=1 in, right=1 in]{geometry}
\usepackage{tikz}
\usetikzlibrary{quantikz2}

\begin{document}

\preprint{APS}

\title{Optimal Continuous- to Discrete-Variable Bipartite Entanglement Conversion}

\author{Pak-Tik Fong}
\email{Email: ptf@sfu.ca}
\author{Ruchir Tullu}
\author{Hoi-Kwan Lau}
\affiliation{%
Department of Physics, Simon Fraser University, Burnaby, British Columbia V5A 1S6, Canada}


\date{\today}

\begin{abstract}

Discrete-variable (DV) entanglement is crucial for numerous quantum applications, yet its deterministic generation in many bosonic systems remains experimentally challenging. In contrast, continuous-variable (CV) entanglement can be produced efficiently. We propose two optimal schemes for converting CV bipartite entanglement into DV entanglement using only local operations and classical communication. The first scheme extracts maximally entangled qubit pairs at the theoretically maximal rate, while the second probabilistically produces a maximally entangled qudit pair with the highest average entanglement. In both schemes, we quantify the optimal performance and identify the measurement operators required for implementation. Notably, using only a sequence of binary measurements, our approach can succeed in a finite number of measurement rounds on average, even though the CV resource is infinite-dimensional. Our schemes improve the feasibility of implementing DV-based quantum technologies on bosonic platforms.

\end{abstract}

\maketitle


\section{Introduction}

A wide range of quantum technologies, including quantum communication \cite{Gisin2007, Ekert1991, PhysRevLett.84.4729}, quantum sensing \cite{RevModPhys.89.035002}, and quantum computing \cite{10.1098/rspa.1985.0070}, are based on discrete-variable (DV) units, like qubits. Since only a small number of basis states is required in each degree of freedom, DV technologies can be implemented with a wide range of quantum systems that contain only a few controllable and stable states, such as defect centers \cite{Katsumi2025}. However, these technologies are also occasionally implemented with bosonic systems, in which each degree of freedom (mode) exhibits effectively infinite levels. For example, microwave resonators can be a robust host of quantum information in circuit quantum electrodynamics (QED) architecture \cite{PhysRevA.69.062320, Wallraff2004}, and the high mobility of photons makes them the dominant carriers of information in quantum communication \cite{Kimble1998, RevModPhys.82.1041, Kimble2008}. 

While DV entanglement, in the form of maximally entangled qubits or qudits, is an essential resource for nearly all quantum technologies, its direct generation on bosonic platforms is usually not trivial. For example, generating photonic DV entanglement usually relies on probabilistic processes \cite{PhysRevA.102.012604, PhysRevResearch.3.043031, PhysRevResearch.6.013073, PhysRevA.79.042326} or requires strong nonlinearities \cite{li2025enhancingmicrowaveopticalbellpairs}. By contrast, continuous-variable (CV) Gaussian entanglement can be generated deterministically by quadratic (in mode operators) interaction, which is readily realizable in most bosonic platforms \cite{limitiation2010, rmp2012, PhysRevLett.96.053602, PhysRevB.76.064305, doi:10.1126/science.abf2998, doi:10.1126/science.1244563, PhysRevLett.131.193601, PhysRevLett.107.113601, 9cpm-kr4h, ptf2024}. 


Nevertheless, maximal CV entanglement is unattainable because it involves infinite energy. Directly applying realistic but non-maximal CV entanglement in quantum technologies introduces unwanted noise \cite{PhysRevLett.80.869,PhysRevLett.89.137903,PhysRevLett.102.120501,limitiation2010}. Furthermore, far fewer quantum technologies have been developed for CV quantum information compared with their DV counterparts. These limitations reduce the practical utility of CV entanglement. Converting Gaussian CV entanglement into DV entanglement therefore provides a possible route to resolve this limitation \cite{Krovi2016, PhysRevApplied.17.034071, PhysRevA.72.034101, Tipsmark:13, PhysRevA.84.012302,SON01082002, PhysRevA.75.032336, PhysRevB.69.214502}. Recent theoretical studies have further explored CV-to-DV entanglement conversion as a means of facilitating scalable entanglement distribution in quantum networks \cite{Zhao_2025,Zhao_2026}.

CV-to-DV entanglement conversion has been extensively implemented in optical systems. The most notable example is the maximal qubit entanglement generated via spontaneous parametric down-conversion (SPDC); it is the entangled single-photon qubit states post-selected from a Gaussian photonic state that is generated by quadratic interaction \cite{Krovi2016, PhysRevApplied.17.034071,PhysRevA.72.034101,Tipsmark:13}. Various improvements to the DV entangled photon source have also been considered by using photon addition and/or subtraction \cite{PhysRevA.84.012302,PhysRevA.86.012328}, and noiseless linear amplification (NLA) \cite{PhysRevA.102.063715}. In hybrid platforms, where both physical qubits and bosonic modes are available, entanglement can be transferred from modes to qubits by interaction and measurement \cite{PhysRevA.75.032336, PhysRevB.69.214502, han2025download}. Nevertheless, the efficiency of these schemes strongly depends on the interactions available in each platform. The fundamental optimality of CV-to-DV entanglement conversion remains unclear.

In this work, we address this question using the framework of entanglement transformation \cite{Nielsen1999,Vidal1999}, which considers the conversion of a single bipartite entangled state into another through local operations and classical communication (LOCC). Our aim is to study the optimal performance of generating maximal DV entanglement from a single copy of a two-mode squeezed vacuum (TMSV) state, which is local unitarily equivalent to any two-mode entangled Gaussian pure state \cite{giedke2003entanglementtransformationspuregaussian}. We note that this framework is distinct from entanglement distillation \cite{PhysRevA.53.2046, QCQI_textbbok}, which aims to extract a collection of maximally entangled states from many weakly entangled copies.

\begin{figure*}
    \centering
    \includegraphics[scale=0.3]{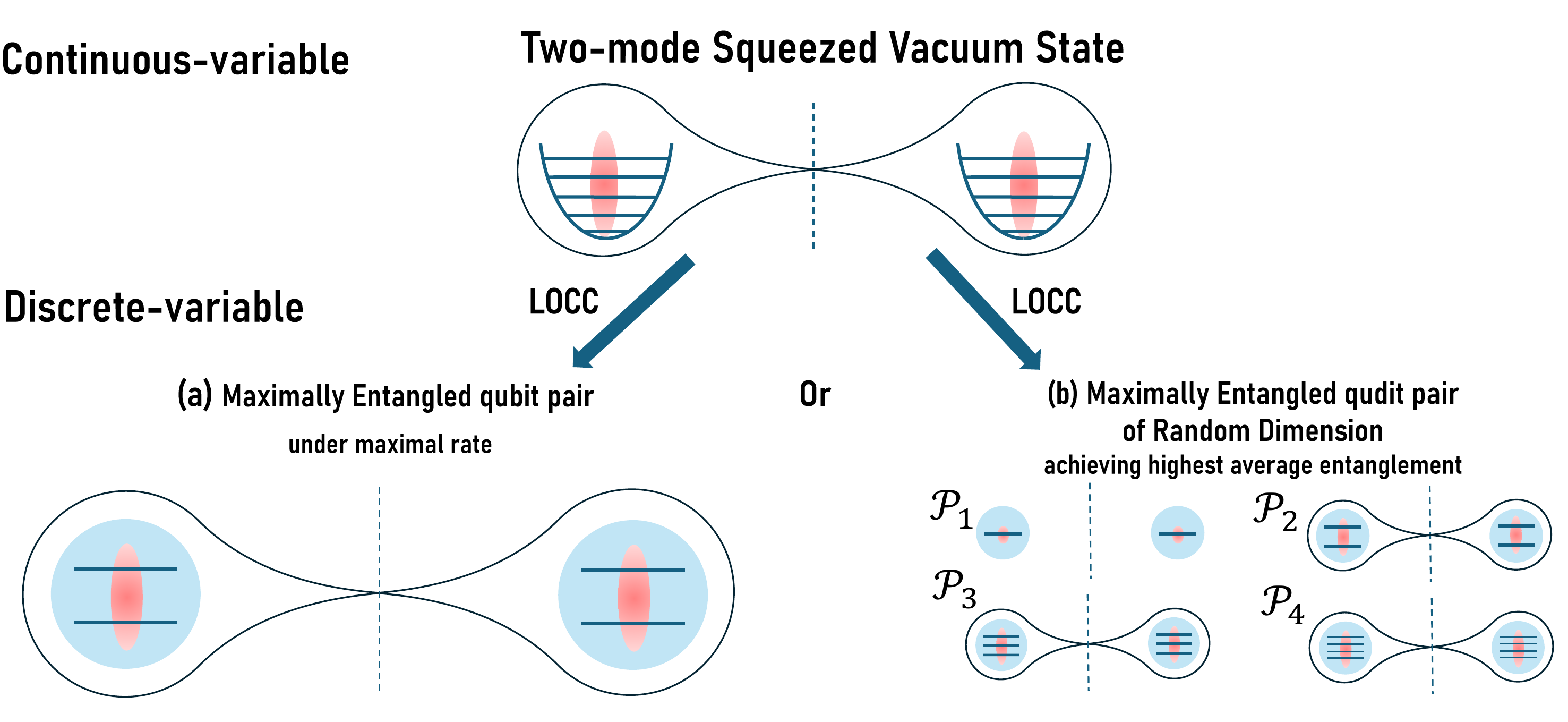}
    \caption{Illustration of CV-to-DV entanglement conversion by transforming a TMSV state into a maximally entangled qudit pair via LOCC. (a) The first scenario considers transforming a TMSV state into a maximally entangled qubit pair with the highest possible rate. (b) In the second scenario, a TMSV is transformed into a maximally entangled qudit pair with random dimension $d$. $\mathcal{P}_d$ is the probability of getting a qudit pair with dimension $d$. The scheme is optimal when the set of probabilities $\{\mathcal{P}_d\}$ gives the average entanglement is the highest.}
    \label{concept}
\end{figure*}

We find that the conversion can be considered optimal in two different and incompatible ways, as illustrated in Fig.~\ref{concept}. The first type of optimality aims to achieve the highest possible success rate of generating a maximally entangled qubit pair. We show that the generation is deterministic whenever the squeezing strength of the TMSV exceeds $7.66$ dB. Below this threshold, we discover the relation between squeezing and the maximum conversion probability, and find that none of the existing schemes achieves the theoretical optimum. We identify the positive operator-valued measures (POVMs) by using Hardy's method of areas \cite{Hardy1999}, which requires at least quadratically fewer POVMs than other known methods \cite{Nielsen1999,PhysRevA.63.062303}.  The second type of conversion aims to produce a maximally entangled qudit pair where the dimension $d$ is randomly determined; it is optimal when the average converted entanglement is maximized \cite{Hardy1999}. We also identify the relation between the amounts of converted entanglement and squeezing, as well as the implementing POVMs.

Because most physical measurement is binary in outcomes, we propose a binary search \cite{binary2008} method to use a sequence of binary measurements to implement the conversion POVMs. Remarkably, we find that the conversion can be accomplished in a finite number of binary measurement rounds on average, even though the CV system is infinite dimensional. We also discuss a possible implementation of the measurements in realistic hybrid qubit-oscillator platforms.

This paper is organized as follows. In Sec.~\ref{scheme1}, we first review the general framework of the entanglement transformation. Then, in Sec.~\ref{sec_TMSV_To_Bell}, we present the optimal conversion scheme for the first scenario, obtaining a maximally entangled qubit pair with the highest probability. In Sec.~\ref{scheme2}, we extend the techniques for the second scenario, achieving the highest average entanglement in the qudits converted. Sec.~\ref{method} discusses the binary search strategy and its efficiency. Then, we discuss the physical implementations of the binary measurements in Sec.~\ref{phyiscal}. Finally, we conclude our work in Sec.~\ref{conclusion}.

\section{Entanglement transformation}\label{scheme1}

We begin by reviewing the framework of the entanglement transformation \cite{Nielsen1999, QCQI_textbbok}. Suppose that two parties, Alice and Bob, share an initial entangled state
\begin{eqnarray}\label{state_A}
    \ket{\Psi_\alpha} = \sum_{n=0}^{\infty} \sqrt{\alpha_n}\ket{u_n}_A\ket{v_n}_B,
\end{eqnarray}
with the Schmidt coefficients $\sqrt{\alpha_n}$. Here, the Schmidt coefficients are arranged in descending order, i.e. $\alpha_n\geq \alpha_{n+1}$. $\{\ket{u_n}_{A}\}$ and $\{\ket{v_n}_{B}\}$ are the Schmidt bases of Alice's and Bob's systems respectively. Our task is to transform this state into a target state with Schmidt coefficients are given by $\sqrt{\beta_n}$, i.e.
\begin{eqnarray}\label{state_B}
    \ket{\Psi_\beta} = \sum_{n=0}^{\infty} \sqrt{\beta_n}\ket{\tilde{u}_n}_A\ket{\tilde{v}_n}_B,
\end{eqnarray}
where $\{\ket{\tilde{u}_n}_A\} $ and $\{\ket{\tilde{v}_n}_B\}$ are the new Schmidt bases which can be transformed from the original ones via local unitaries. 

As discovered by Nielsen \cite{Nielsen1999}, a deterministic transformation via LOCC is possible if and only if the squared Schmidt coefficients of the target state majorize those of the initial state, i.e. 
\begin{eqnarray}\label{cumulative probabilities}
    \sum_{n=l}^\infty \alpha_n \geq \sum_{n=l}^\infty \beta_n,
\end{eqnarray}
for all integer $l\in[0,\infty)$. While this condition implies that the target state cannot have more entanglement than the initial state, we note that the amount of entanglement is not the only factor that determines the possibility of deterministic transformation. It is also determined by the detailed distribution of the Schmidt coefficients \cite{majorization}. 

When the majorization condition \eqref{cumulative probabilities} is not satisfied, the transformation cannot be performed deterministically but only probabilistically. Vidal showed that the maximum success probability is determined by the minimum ratio of the cumulative probabilities \cite{Vidal1999}, i.e.
\begin{eqnarray}\label{Pmax}
    P_{\text{max}} = \min_{l\in[0,\infty)} \frac{\sum_{n=l}^\infty \alpha_n}{\sum_{n=l}^\infty \beta_n},
\end{eqnarray}
where the minimization is restricted to nonzero denominators.

\section{Converting TMSV to Maximally entangled qubit Pair}\label{sec_TMSV_To_Bell}

TMSV states can be generated from vacuum by using parametric amplification interaction that is widely available in bosonic platforms \cite{limitiation2010, rmp2012, PhysRevLett.96.053602, PhysRevB.76.064305, doi:10.1126/science.abf2998, doi:10.1126/science.1244563, PhysRevLett.131.193601, PhysRevLett.107.113601, 9cpm-kr4h}. It is also known that any pure two-mode Gaussian state, which is the CV state that can be produced from vacuum by applying general quadratic interaction, can be transformed to TMSV by using only local Gaussian unitaries \cite{giedke2003entanglementtransformationspuregaussian}. TMSV states have wide utilities in CV quantum technologies because it exhibits quantum correlations between the continuous quadratures of two modes, i.e. the correlated fluctuations of $q$-quadratures and anti-correlated fluctuations of $p$-quadratures are both below the shot-noise limit, $\braket{(\hat{q}_A-\hat{q}_B)^2} =\braket{(\hat{p}_A+\hat{p}_B)^2} = e^{-2r} <1$. Here $\hat{q}$ and $\hat{p}$ are respectively the $q$- and $p$-quadrature operators; subscripts $A$ and $B$ denote the modes of Alice and Bob respectively; $r$ is the squeezing parameter. The infinitely squeezed TMSV state, i.e. $r\to \infty$, is also known as Einstein-Podolsky-Rosen (EPR) state because it resembles the quantum state in the seminal thought experiment \cite{PhysRev.47.777}.

To study the CV-to-DV entanglement conversion, we express the TMSV state in Schmidt decomposition \cite{rmp2012},
\begin{eqnarray}
    \ket{\text{TMSV}} = \sqrt{1-\lambda^2}\sum_{n=0}^\infty \lambda^n \ket{n}_A\ket{n}_B, \label{TMSV}
\end{eqnarray}
where $\lambda \equiv \tanh r$ and $\ket{n}$ is the $n$-boson Fock state. The squared-Schmidt coefficients are given by
\begin{eqnarray}\label{alpha}
   \alpha_n&=&(1-\lambda^2)\lambda^{2n} \quad \text{for} \quad n\in[0,\infty).
\end{eqnarray}
Our target is to obtain a maximally entangled qubit pair, i.e. the state with the squared-Schmidt coefficients
\begin{eqnarray}\label{beta}
     \beta_n &\equiv &
    \begin{cases}
        1/2 \quad & \text{for} \quad n=0,1\\
        0 \quad &\text{otherwise}
    \end{cases}.
\end{eqnarray}
We note that the TMSV state possesses infinitely many nonzero Schmidt coefficients, whereas the DV maximally entangled qubit pair has only two.

\subsection{Optimal Conversion Rate}\label{sec_optimal_rate}

By using the majorization condition \eqref{cumulative probabilities}, we find that a maximally entangled qubit pair can be deterministically converted from a TMSV only when $\lambda \geq 1/\sqrt{2}$, which corresponds to a squeezing strength greater than $7.66$ dB. We refer to this as the {\it threshold squeezing} strength. Notably, the amount of entanglement possessed by a threshold TMSV is equal to $2$ ebits. It shows that the possibility of entanglement transformation is not determined solely by the amount of entanglement contained in the initial and target states. Generally, there is a price to pay to obtain an entangled state with the desired structure, i.e. the desired distribution of Schmidt coefficients.

When the squeezing strength is below the threshold value, $\lambda<1/\sqrt{2}$, the conversion cannot succeed with certainty. By using Eq.~\eqref{Pmax}, we find that the maximum success probability is given by $2\lambda^2$. Therefore, the maximal success probability of the entanglement conversion is summarized as
\begin{eqnarray}\label{max_P}
    P_{\text{max}} = \text{min}(1,2\lambda^2).
\end{eqnarray}

As mentioned, CV-to-DV entanglement conversion plays a significant role in bosonic quantum technologies, so various implementations have been proposed for different platforms. It would be interesting to know how efficient the existing schemes are, when comparing to the theoretical maximum Eq.~\eqref{max_P}. We define the conversion rate as the success probability for obtaining one copy of maximally entangled qubit pair per each TMSV state resource. 

\begin{figure}
    \centering
    \includegraphics[scale=0.26]{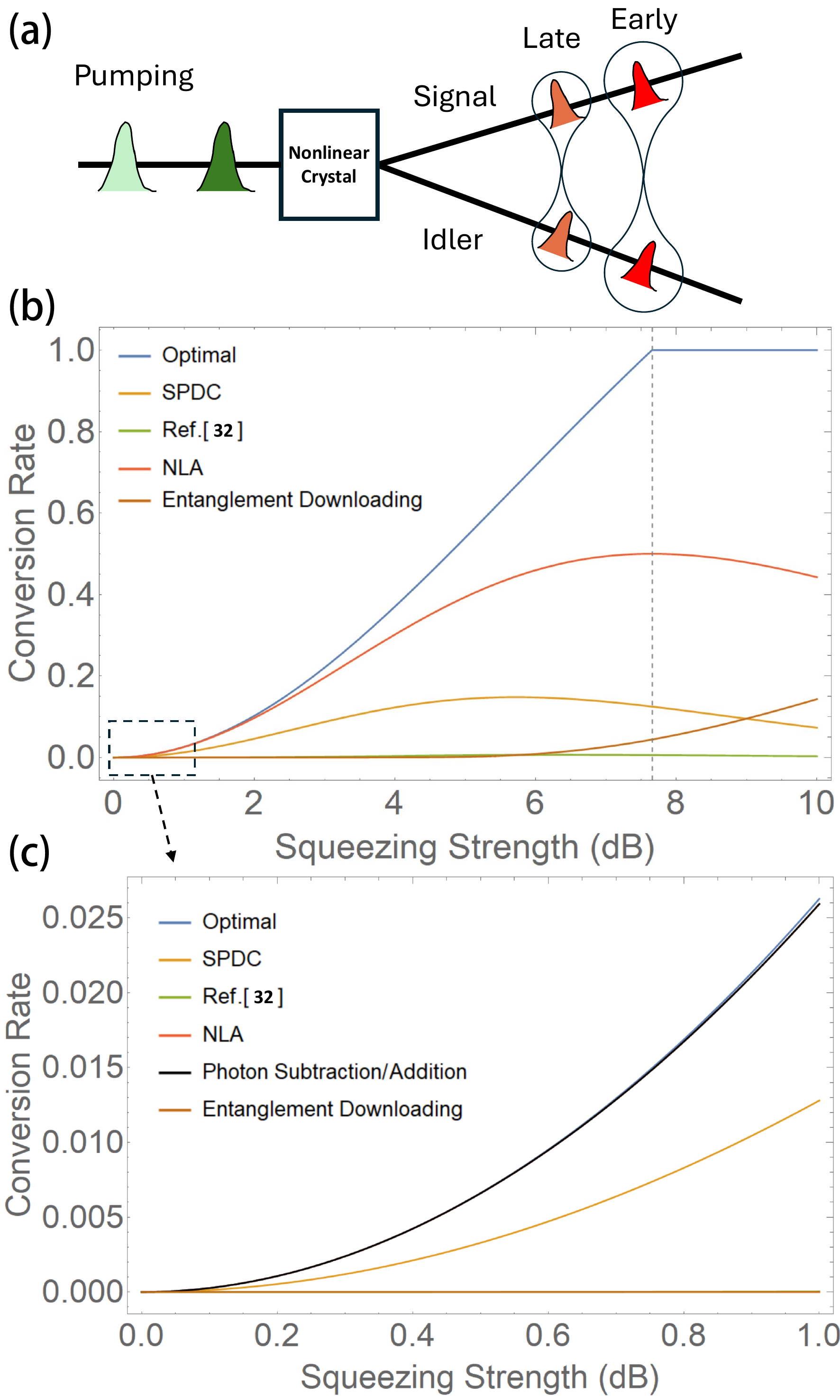}
    \caption{(a) The illustration of the probabilistic generation of a maximally entangled single-photon time-bin qubit pair from the SPDC photon source. (b) The conversion rates of our optimal scheme and several existing schemes. The vertical gray-dashed line indicates the threshold squeezing strength. The dashed box indicates the weak-squeezing regime. (c) The zoom-in of the weak-squeezing regime. In this regime, we also include the curve corresponding to the scheme based on a superposition of photon addition and subtraction \cite{PhysRevA.84.012302,PhysRevA.86.012328}. Unlike other schemes shown in (b), this method produces a state that approximates a maximally entangled qubit pair with high fidelity only in the weak-squeezing limit. Details can be found in Appendix~\ref{app_conversion_rate}.}
    \label{optimal_rate}
\end{figure}

A widely used entangled photon source in linear quantum optics experiments is SPDC, in which a nonlinear crystal is pumped by a classical laser. The pump activates a parametric interaction that will generate a TMSV state \cite{PhysRevA.61.042304} between a degree of freedom in the signal and another in the idler. The choice of degrees of freedom varies across implementations, they could be, e.g. time bins or polarizations \cite{RevModPhys.84.777}. Generally, the pump is weak in the sense that the state is dominated by the vacuum and single-photon-pair components. In experiments, two pairs of TMSV are usually generated in different degrees of freedom. As an example, we illustrate in Fig.~\ref{optimal_rate} (a) the time-bin entanglement source, where two pump pulses are applied sequentially and produce a TMSV between each pair of early and late time bins \cite{PhysRevLett.66.1142, Xavier2025}.
The total state is given by
\begin{eqnarray}\label{spdc}
\ket{\text{TMSV}}^{\otimes 2} &=& (1-\lambda^2)\Big(\ket{00}_s \ket{00}_i  \nonumber\\
&+&  \lambda \ket{01}_s \ket{01}_i + \lambda \ket{10}_s \ket{10}_i+O(\lambda^2)\Big),\nonumber\\
\end{eqnarray}
where the subscripts $s$ and $i$ respectively denote the signal and idler modes; the left and right indices in each ket respectively denote the photon number in the late and early time bins. Experiments usually proceed with Alice and Bob, respectively, collect the signal and idler modes, manipulate the collected modes, and subject them to photon detection. By post-selecting the events that both Alice and Bob record a single-photon click, the observations are originated from the equal superposition of early- and late-time photon pairs (i.e. second and third terms in Eq.~\eqref{spdc}), which corresponds to a maximally entangled single-photon qubit pair \cite{PhysRevLett.98.190503, Krovi2016}. Overall, with perfect collection efficiency, SPDC photon source will generate a maximally entangled qubit pair from two TMSV. 

Apart from SPDC, other optical protocols have been proposed to generate maximally entangled qubit pairs from TMSV through number-resolving detections \cite{PhysRevA.72.034101}, superposition of photon addition and subtraction \cite{PhysRevA.84.012302,PhysRevA.86.012328}, and NLA \cite{PhysRevA.102.063715}. For hybrid qubit-oscillator systems, entanglement between two qubits can be established through interacting with the entangled oscillators \cite{PhysRevA.75.032336, PhysRevB.69.214502}. A recent scheme suggests that entanglement can be downloaded from a two-mode CV cluster state, which can be converted from TMSV by local unitaries, to two physical qubits \cite{han2025download}. 

In Figs.~\ref{optimal_rate} (b) and (c), we compare the conversion rates of these established methods with the theoretical optimum. Details of these methods can be found in Appendix~\ref{app_conversion_rate}. We find that no known scheme is optimal in all range of squeezing. Special interest is given to the weak-squeezing regime that most optical schemes are operating at. The conversion rate of the SPDC source \cite{PhysRevLett.66.1142,PhysRevLett.98.190503, Krovi2016} is found to be roughly half of $P_{\text{max}}$ when $\lambda \ll 1$. The sub-optimality originates from the post-selection process based on photon measurement: it collapses the superposition, and thus breaks the entanglement, between vacuum and the non-zero-photon states. Nevertheless, given the severe restriction of linear optics, particularly all passive optical elements must preserve the photon number, the mere half performance deficient is surprisingly minor. For other optical schemes, we find that superposition of photon addition and subtraction \cite{PhysRevA.75.032336, PhysRevB.69.214502} and NLA \cite{PhysRevA.102.063715} are almost optimal in the weak-squeezing regime. It is because in this regime their projectors coincides with the dominant POVM in the optimal scheme, which we will discuss further in the next subsection. For hybrid systems, the conversion rate of the entanglement downloading \cite{han2025download} is low. It is understandable because the scheme focuses on preserving the structure of many-body entanglement, while the two-body efficiency is not a priority. 

\subsection{POVMs for Optimal Scheme}\label{beyond}

To implement entanglement transformation, Nielsen \cite{Nielsen1999, QCQI_textbbok} suggests a protocol that involves sequential measurements and local transformation by both Alice and Bob. The method is later simplified by Hardy by using the "method of areas" \cite{Hardy1999} and by Jensen and Schack by using the Birkhoff–von Neumann theorem \cite{PhysRevA.63.062303}. These schemes can be characterized by the POVMs corresponding to all possible measurement outcomes obtained during the processes. Since measurement is the most time-consuming and noisy element in many physical platforms, we would like to reduce the number of measurements involved in a CV-to-DV entanglement conversion. It is thus interesting to know the number of POVMs required by each scheme; as will be discussed in Sec.~\ref{method}, the number of POVMs is related to the required rounds of binary measurement.

\begin{figure}
    \centering
    \includegraphics[scale=0.26]{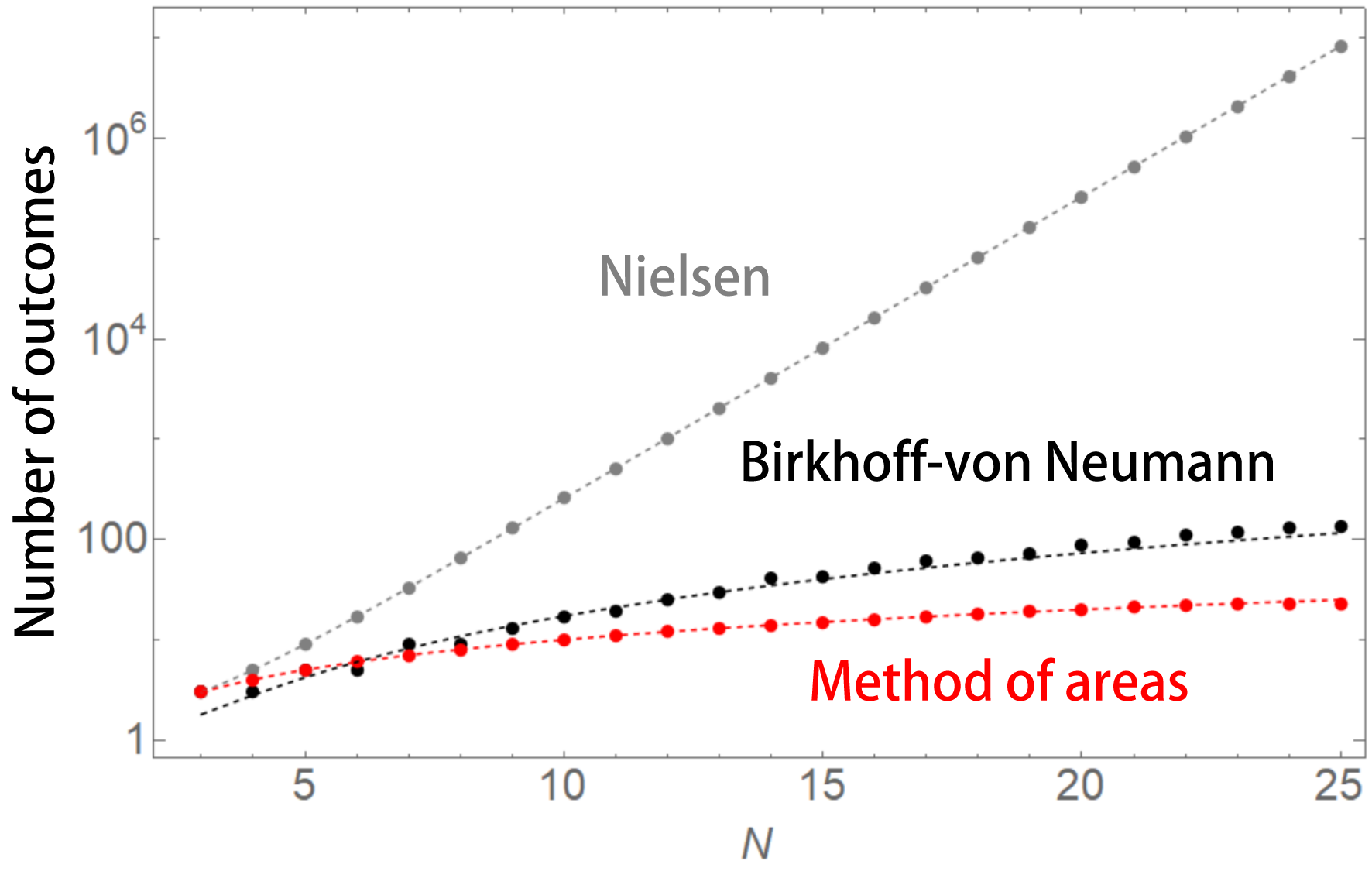}
    \caption{The number of POVMs yielded by different entanglement transformation schemes against the boson number truncation $N$ of a TMSV. The presented data corresponds to $\lambda = 0.5$, but the behavior is typical for other values of $\lambda$. Nielsen's scheme, Birkhoff-von Neumann's algorithm and method of areas require $2^{N-2}+1$, $ (N-1)^2/5+1$ (empirical scaling fit) and $N$ POVMs, respectively.}
    \label{number_of_operators}
\end{figure}

A TMSV state contains infinitely many nonzero Schmidt coefficients. In practice, we can only address a finite number of them, so we truncate the state in Eq.~\eqref{TMSV} to consider only the $N$ largest Schmidt coefficients, i.e. up to $N-1$ excitations in each mode. In Fig.~\ref{number_of_operators}, we compare the number of outcomes required in different CV-to-DV entanglement conversion schemes. We find that Nielsen's method \cite{Nielsen1999, QCQI_textbbok} almost always removes one Schmidt coefficient in each round of binary measurement, so the whole process takes $\mathcal{O}(N)$ rounds of measurement, which consists of $\mathcal{O}(2^N)$ measurement outcomes. Detailed discussion can be found in Appendix \ref{app_NMethod}. Jensen and Schack's method \cite{PhysRevA.63.062303} relies on decomposing the double-stochastic matrix that maps the vector $\{\beta_n\}$ to $\{\alpha_n\}$ into permutations. Since the double-stochastic matrix is $N\times N$ in dimension and each permutation represents one POVM, there will be $\mathcal{O}(N^2)$ POVMs. Finally, by using Hardy's method, we find that CV-to-DV conversion can be completed with only $\mathcal{O}(N)$ POVMs \cite{Hardy1999}. This represents at least quadratic reduction of POVMs when compared to other schemes. In the following, we illustrate the application of Hardy's method in CV-to-DV conversion.

\subsection{Hardy's Method of Areas}

\begin{figure*}
    \centering
    \includegraphics[scale=0.3]{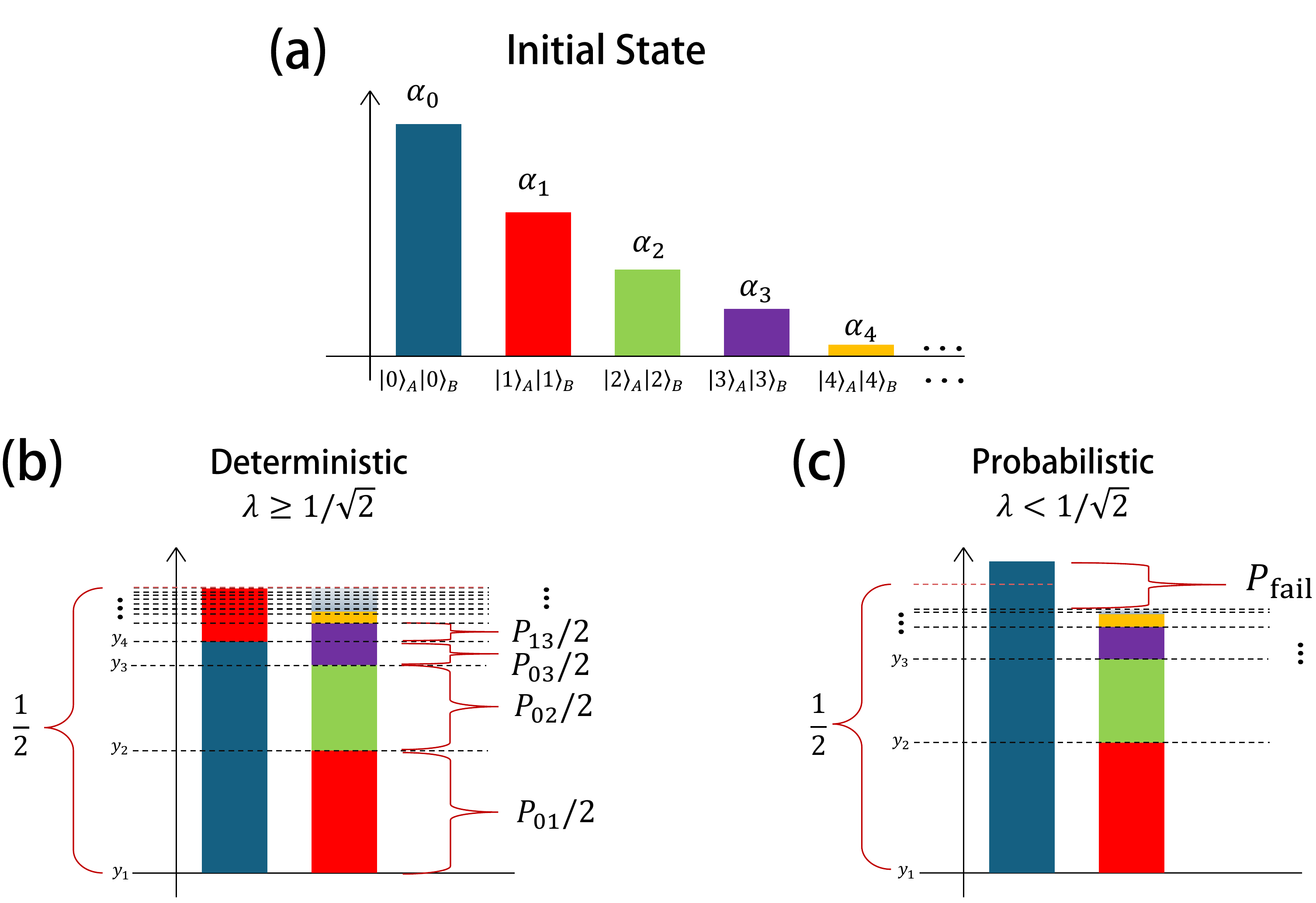}
    \caption{Illustration of Hardy's method of areas. (a) The chart represents the initial TMSV state. Every column is painted in a different colour, and the area covered by each column is given by the squared Schmidt coefficients $\{\alpha_n\}$. (b) The transformed chart in the deterministic regime. The two columns have the same height $\beta_0=\beta_1=1/2$, so the chart coincides with that represents the target state. The boundaries of adjacent colours are labeled by $\{y_l\}$. The probability of obtaining $\ket{\text{Bell}_{nm}}$, $P_{nm}$, is determined by the area covered by the segment of columns between two adjacent boundaries, and the corresponding Fock states involved, $\{n,m\}$, are determined by the colours of the segment.
    (c) The transformed chart in the probabilistic regime. The heights of the first and second columns are respectively $\alpha_0$ and $\sum_{n=1}^\infty\alpha_n$. Since $\alpha_0 > 1/2$, there is a segment consisting of only the first but not the second column. This segment, whose height and covered area are both $P_{\text{fail}}$, corresponds to the failed conversion. }
    \label{method_area}
\end{figure*}

Hardy’s method of areas expresses the squared Schmidt coefficients as a column chart ordered by decreasing height, where each column is unit-width, painted with a distinct colour, and has an area equal to the corresponding Schmidt coefficient. As an example, Fig.~\ref{method_area} (a) shows the chart of a TMSV. Hardy showed that entanglement transformation is equivalent to dividing the columns into patches and stacking the patches onto the columns to the left without repeating any colours at any horizontal level, \cite{Hardy1999}. See Appendix~\ref{app_method_area} for the details. If this process can form the chart of the target state, e.g. Fig.~\ref{method_area} (b) for a maximally entangled qubit pair, the scheme is deterministic. Otherwise, a transformed chart should be constructed to fit in a compressed version of the target chart. Exploiting all possible chart transformations, the one that fits in a least compressed target chart, e.g. Fig.~\ref{method_area} (c), deduces the maximum probability of successful transformation. The pattern of the transformed chart also determines the measurement operators that implement such transformation \cite{Hardy1999}.

To generate a maximal entangled qubit pair, the target chart consists of two equal-height columns. The columns can be divided into segments, such that in each segment there is only one colour in each column.
We indicate the segment boundaries by dashed lines in Fig.~\ref{method_area} (b) and (c). 
According to Hardy's method, the colours of each segment tell the initial Schmidt states that will be involved in the target state after transformation, and the total area covered by the segments will be the probability of obtaining such outcome. In our case, because the Schmidt bases of TMSV are paired Fock states, the target states will be equal superpositions of Fock states, i.e.
\begin{eqnarray} \label{eq_bell}
    \ket{\text{Bell}_{nm}} = \frac{1}{\sqrt{2}}\left(\ket{n}_A\ket{n}_B +  \ket{m}_A\ket{m}_B\right),
\end{eqnarray}
where $n\neq m$. 

When the squeezing is above threshold, we can construct a transformed chart that consists of two columns with equal height as $1/2$, see Fig.~\ref{method_area} (b). Because it coincides with the target chart, the transformation is deterministic.  When the squeezing is below threshold, the first initial column has height $1- \lambda^2 > 1/2$, so it is higher than all the other columns combined. After transformation, the two columns thus have different height, so the conversion is not deterministic. Then there will be a segment in the first column that cannot pair with any patch in the second column. It corresponds to the failed conversion that the outcome becomes a separable state $\ket{0}_A\ket{0}_B$. 

The probability of getting a particular outcome is determined by the total area covered by the corresponding segments, which is the sum of the heights of the segments multiplied with the unit width. 
When $\lambda\geq 1/\sqrt{2}$, the deterministic regime, all segments are paired, so there is no failed conversion, i.e. the fail probability $P_{\text{fail}}=1-P_{\text{max}}=0$. 
We note that in the deterministic regime there is no unique way to construct the transformed chart.
For example, one can fill up the first column by using the patch cut from the second or any other columns.
In this work, as illustrated in Fig.~\ref{method_area} (b), we consider a specific construction that fills the first transformed column with the patches from the first $M$ initial columns, where the sum of areas covered by these columns is just larger than $1/2$, i.e. $\sum_{j=0}^{M-1} \alpha_j \geq 1/2 > \sum_{j=0}^{M-2} \alpha_j$. Then, the second transformed column consists of a patch of the $M$-th initial column with height $\sum_{j=0}^{M-1}\alpha_j-1/2$ and all the remaining initial columns. The segment boundaries are located at the colour boundaries of the first columns, $\{\sum_{i=0}^n\alpha_i\}$ for $n < M-1$, as well as those of the second columns, $\{\sum_{i=0}^n\alpha_i -1/2\}$ for $n \geq M-1$. We rank them in ascending order, $\{y_0=0,y_1,y_2,y_3,...\}$. The probability of the $l$-th outcome will be given by $2(y_{l} - y_{l-1})$ for $l \in [1,\infty)$, and the corresponding Fock states involved in that outcome can be deduced from the colours of columns between the boundaries $y_{l}$ and $y_{l-1}$.

For $\lambda < 1/\sqrt{2}$, the probabilistic regime, the fail probability $P_{\text{fail}}$ is non-zero and given by the height of that unpaired part. It can be minimized by forming the first transformed column with only the first column of the initial chart, while the second transformed column consists of all the remaining initial columns, as illustrated in Fig.~\ref{method_area} (c). This arrangement gives the maximum success probability, because moving any patch to the first column will only reduce the transformation success probability.
The fail probability can be easily found as $P_{\text{fail}}=1-P_{\text{max}}=1-2\lambda^2$. In this case, the boundaries of segments are simply given by $y_l=\sum_{j=1}^{l}\alpha_j$ for $l\in[1,\infty)$. The segment between the boundaries $y_l$ and $y_{l-1}$ corresponds to the $\ket{0}_A\ket{0}_B$ for the first column and $\ket{l}_A\ket{l}_B$ for the second column. Hence, the outcomes are $\ket{\text{Bell}_{0l}}$ and the corresponding probabilities are given by $P_{0l} = 2\lambda^{2l}(1-\lambda^2)$ for $l\geq 1$.

Implementing the entanglement transformation deduced from Hardy's method requires only local measurements on one party, taken here to be Alice without loss of generality. The corresponding measurement operators $\hat{M}_{nm}$ and $\hat{M}_{\text{fail}}$ should respectively project the TMSV to the target states, $\ket{\text{Bell}_{nm}}$ and $\ket{0}_A\ket{0}_B$, with the probabilities $P_{nm}$ and $P_{\text{fail}}$ specified by the Hardy's method, i.e.
\begin{eqnarray}\label{projection}
    \hat{M}_{nm}\ket{\text{TMSV}} &=& \sqrt{P_{nm}}\ket{\text{Bell}_{nm}},\nonumber\\
    \hat{M}_{\text{fail}} \ket{\text{TMSV}} &=& \sqrt{P_{\text{fail}}}\ket{0}_A\ket{0}_B.
\end{eqnarray}
They can be easily found as
\begin{eqnarray}
\hat{M}_{nm} &=& \sqrt{\frac{P_{nm}}{2(1-\lambda^2)}}\nonumber\\&\times&\left(\lambda^{-n} \ket{n}_A\bra{n}+\lambda^{-m}\ket{m}_A\bra{m}\right),\nonumber\\
     \hat{M}_{\text{fail}} &=& \sqrt{\frac{P_{\text{fail}}}{1-\lambda^2}} \ket{0}_A\bra{0},
\end{eqnarray}
which are diagonal in the Fock basis. It is easy to check that these POVMs form a complete set in both deterministic and probabilistic regimes, i.e.
\begin{eqnarray}\label{normalization}
     \hat{M}_{\text{fail}}^\dag\hat{M}_{\text{fail}} + \sum_{n}\sum_{m}\hat{M}_{nm}^\dag \hat{M}_{nm} = \sum_{n=0}^{\infty}\ket{n}_A\bra{n}.\nonumber\\
\end{eqnarray}

The measurement corresponding to these POVMs can be realized by a single measurement process that has infinite outcomes \cite{PhysRevA.63.062303}, but this is usually difficult to implement in practice. Alternatively, we will discuss in Sec.~\ref{method} a measurement scheme that requires only a sequence of binary measurements.

\section{Converting TMSV to maximally entangled qudit pair with random dimension}\label{scheme2}

Entanglement of a TMSV state increases with squeezing, so it is expected that more DV entanglement can be extracted from a stronger squeezed TMSV state. However, the average number of maximally entangled qubit pairs obtained from a TMSV plateaus at $7.66$ dB. One could straightforwardly extend the entanglement transformation to create maximally entangled qudit pairs, i.e. the state with the squared-Schmidt coefficients
\begin{eqnarray}
\beta_n=\begin{cases}
        1/d \quad & \text{for} \quad n=0,1,...d-1\\
        0 \quad &\text{otherwise}
    \end{cases},
\end{eqnarray}
but such schemes will also plateau when the squeezing is sufficiently high that the transformation becomes deterministic. Nevertheless, maximal DV entanglement can also be obtained in other forms: entanglement concentration protocol is known to convert any bipartite entangled state to a maximally entangled qudit pair, where the dimension $d$ is random \cite{PhysRevA.53.2046, generalized_nielsen1999}. The probabilities $\mathcal{P}_d$ of obtaining each $d$-dimensional entangled qudit pair depend on the local operations. For the optimal scheme that generates the highest average entanglement, the probabilities are known to be $\mathcal{P}_d = d\left(\alpha_{d-1}-\alpha_d\right)$ \cite{Hardy1999}. For CV-to-DV entanglement conversion, it is given by 
\begin{eqnarray}
    \mathcal{P}_d &=&  d\lambda^{2d-2}(1-\lambda^2)^2.
\end{eqnarray}
The resulting average entanglement is 
\begin{eqnarray}\label{s_avg}
    S_{\text{avg}} &=& \sum_{d=1}^\infty \mathcal{P}_d  S_d \nonumber\\
    &=&-\frac{(1-\lambda^2)^2 }{\lambda^{2} \ln2}\frac{\partial \text{Li}_s(\lambda^2)}{\partial s}|_{s=-1},
\end{eqnarray}
where $S_d=\log_2d$ is the entanglement entropy of the $d$-dimensional maximally entangled qudit pair \cite{10.5555/2011706.2011707} and $\text{Li}_s(z)$ is a polylogarithm of order $s$ \cite{polylogarithm_book}.

\begin{figure}
    \centering
    \includegraphics[scale=0.37]{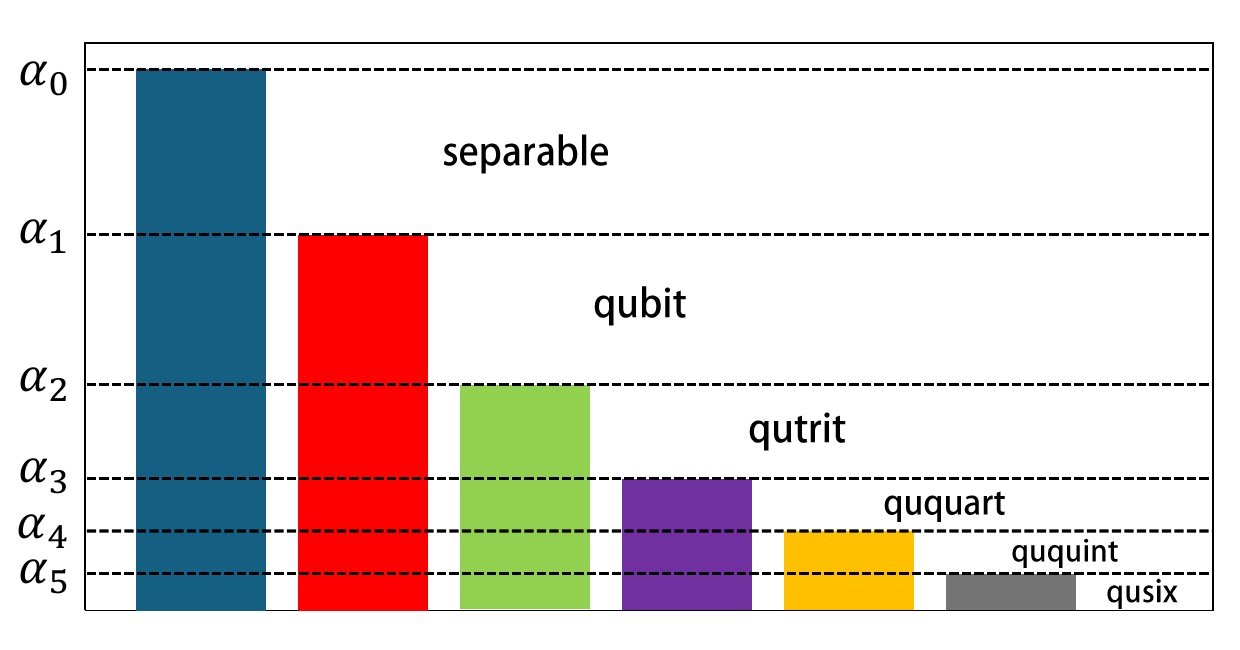}
    \caption{Illustration of the method of areas for the optimal random qudit scheme. The segment boundaries (dashed) are located at the heights of the columns. The probability of obtaining a qudit pair with dimension $d$ is given by the total area of columns between boundaries at $\alpha_{d-1}$ and $\alpha_d$, and the corresponding Fock states involved are determined by the columns included in that segment.}
    \label{method_area_class}
\end{figure}

The composition of the obtained entangled qudit pair, and hence the corresponding POVMs, can be learned from the method of areas, as illustrated in Fig.~\ref{method_area_class}. 
Hardy found that the maximum average entanglement is obtained without transforming the chart. In this situation, the segment boundaries will be given by the heights of all columns, i.e. $\{\alpha_n\}$. The maximally entangled DV state of dimension $d$ is specified by the segment that involves $d$ columns, which is located between the boundaries at $\alpha_{d-1}$ and $\alpha_d$. The entangled qudit pair obtained will be an equal superposition of the first $d$ components in the Schmidt decomposition, i.e. the equal superposition of the first $d$ Fock states in both modes,
\begin{eqnarray}
     \ket{\varphi_d}  = \frac{1}{\sqrt{d}}\sum_{n=0}^{d-1}\ket{n}_A\ket{n}_B.
\end{eqnarray}
Considering the above state can be generated by the POVM $\hat{\mathcal{M}}_d$, 
\begin{eqnarray}
    \hat{\mathcal{M}}_d\ket{\text{TMSV}_\lambda} = \sqrt{\mathcal{P}_d}\ket{\varphi_d},
\end{eqnarray}
the measurement operators can be found as
\begin{eqnarray}\label{scheme2_M}
    \hat{\mathcal{M}}_d &\equiv& \lambda^{d-1}\sqrt{1-\lambda^{2}}\sum^{d-1}_{n=0} \lambda^{-n}\ket{n}_A\bra{n},
\end{eqnarray}
for $d=1,2,...,\infty$. They satisfy the completeness condition $\sum_{d=1}^{\infty}\hat{\mathcal{M}}_d^\dag\hat{\mathcal{M}}_d = \hat{I}$.

\begin{figure}
    \centering
    \includegraphics[scale=0.26]{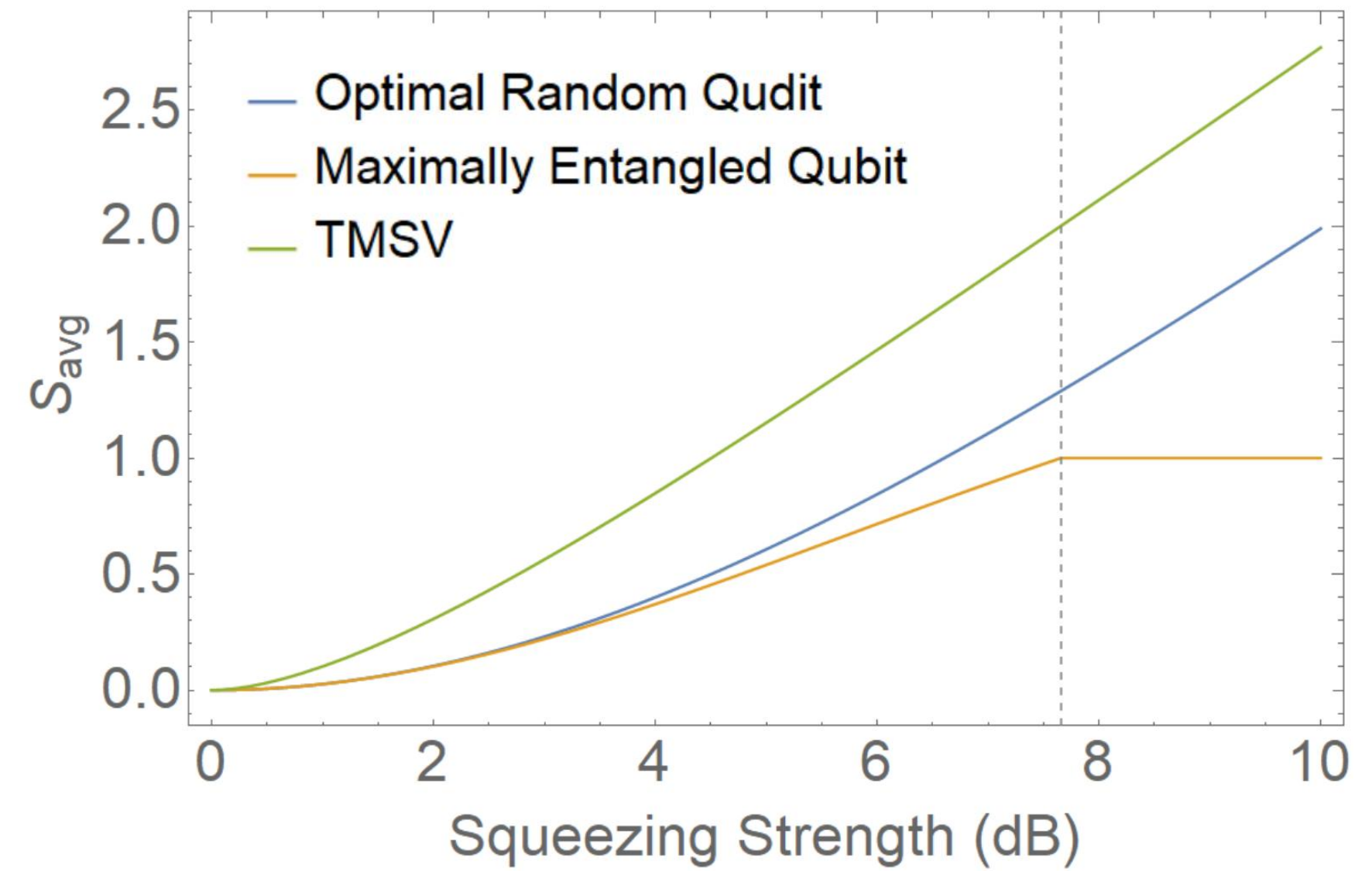}
    \caption{Entanglement of the original TMSV state and the converted maximally entangled DV states versus squeezing strength. }
    \label{average}
\end{figure}

In Fig.~\ref{average}, we compare the entanglement entropy of a TMSV state with the average DV entanglement obtained under the optimal schemes in Secs.~\ref{sec_TMSV_To_Bell} and \ref{scheme2}. The entanglement generated by the random qudit scheme increases monotonically with the squeezing strength, in stark contrast to the maximal qubit scheme that saturates. Moreover, as squeezing increases, the gap between the entanglement of the initial TMSV state and the average DV entanglement obtained from the optimal random qudit scheme approaches a constant value, $\gamma/\text{ln(2)} \approx 0.832746$, where $\gamma$ is the Euler–Mascheroni constant \cite{euler_book}. The derivation of this quantity is given in Appendix \ref{app_constant}.

\section{Realizing POVMs by Binary Search}\label{method}

Since the measurement in most quantum platforms is binary, instead of implementing a many-outcome measurement \cite{PhysRevLett.73.58}, we propose to realize the POVMs by binary search, i.e. a sequence of binary measurements \cite{wang2006, binary2008,PhysRevB.95.134501,PRXQuantum.5.030315}.  

For convenience, hereafter, we will relabel the required POVMs and the corresponding outcomes for any scheme as $\{\hat{E}_1,\hat{E}_2,...,\hat{E}_\infty\}$ and $\{E_1,E_2,...,E_\infty\}$ respectively; the POVMs are arranged in descending order of the probabilities, i.e. $P(E_1)\geq P(E_2)\geq ...\geq P(E_\infty)$. 

\subsection{One-Outcome-per-Round Binary Search}\label{sec_ocpr}

\begin{figure}
    \centering
    \includegraphics[scale=0.37]{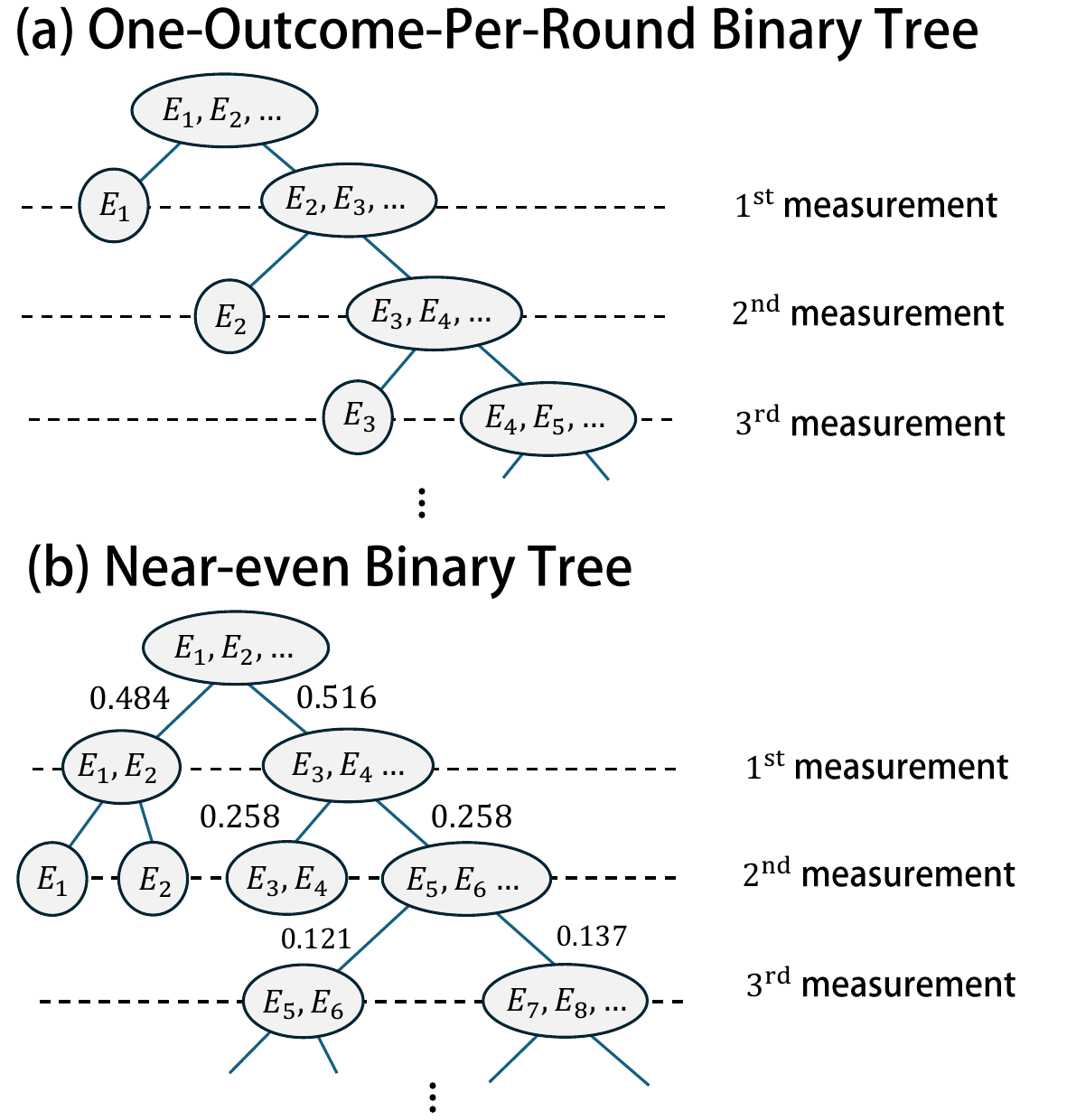}
    \caption{Illustration of binary search trees. (a) The simple one-outcome-per-round binary search, which determines only one outcome in each round of binary measurement. (b) The near-even binary search for the case of $\lambda=0.8$. In each round of binary measurement, the outcomes are divided into two groups according to Eq.~\eqref{eq_prob_div}. The probabilities of the binary outcomes are labeled beside the branches leading to the corresponding group. }
    \label{tree2}
\end{figure}

A straightforward implementation is to distinguish one outcome in each round of measurement. Following the general construction of sequential binary measurements \cite{binary2008}, the binary POVMs implemented in the $m$-th round are 
\begin{eqnarray}\label{one_out}
\left\{
\hat E_m\hat R^{-1/2}, \,
\left(\sum_{i=m+1}^{\infty}\hat E_i^\dagger\hat E_i\right)^{1/2}\hat R^{-1/2} + \hat{Q}\right\},
\end{eqnarray}
where $\hat R=\sum_{i=m}^{\infty}\hat E_i^\dagger\hat E_i$ is the residual POVM effect, $\hat{R}^{-1/2}$ denotes the Moore–Penrose pseudoinverse of $\hat{R}^{1/2}$ and $\hat{Q}$ is the identity operator in null space of $\hat R$. We remark that the products involving the pseudoinverse are understood as their bounded extensions, whose existence follows from Douglas' factorization theorem \cite{Douglas1966}. These operators satisfy the completeness condition. Their construction guarantees that the state is projected to $\hat{E}_m \ket{\text{TMSV}}$ when the outcome $E_m$ is obtained.  We illustrate the idea with a binary tree in Fig.~\ref{tree2} (a). The binary measurements are performed sequentially until an outcome is distinguished. 

\subsection{Near-Even Binary Search}

The above one-outcome-per-round (Sec.~\ref{sec_ocpr}) scheme is not the only way to construct a binary tree; alternative construction can further reduce the measurement rounds. We propose a near-even binary search, illustrated in Fig.~\ref{tree2} (b), in which the two binary outcomes in each round are chosen to have probabilities as close as possible.

Our intuition is that our task of distinguishing a particular outcome is equivalent to revealing the identity $E_m$ of a random variable that follows the probability $P(E_m)$.  Every binary measurement provide at most $1$ bit of information about the random variable, and the amount of information is higher when the probabilities of the two outcomes are closer to each other.  Therefore, by grouping the possible outcomes into two sets that have close probabilities in every binary measurement, we increase the information gained about the random variable in each round, thus reducing the required number of measurement rounds.  

Nevertheless, grouping numbers into two sets that have minimal sum difference is known to be a hard problem \cite{10.3389/fphy.2014.00005}. Here we aim to use only a simple routine to reduce the probability difference between two groups of outcomes, instead of looking for a grouping that minimizes the difference. Consider a task that aims to distinguish an outcome from $M$ possible ones $\{E_1, ..., E_M\}$. We divide them into two groups $B_0=\{E_1, ..., E_Y\}$ and $B_1=\{E_{Y+1}, ..., E_M\}$, where $Y$ is the largest number that the first group has less total probability than the second, i.e.
\begin{equation}\label{eq_prob_div}
\sum_{i=1}^{Y+1} P(E_i) > \sum_{i=1}^M P(E_i)/2 \geq \sum_{i=1}^Y P(E_i).
\end{equation}
However, if the most probable outcome already exceeds half of the total probability, we set $Y=1$.
A binary measurement can be constructed as $\{\hat{B}_0,\hat{B}_1\}$, where
\begin{eqnarray}
    \hat{B}_{0} &=&  \left(\sum_{i=1}^Y \hat{E}_i^\dag\hat{E}_i\right)^{1/2}\hat{R}^{-1/2},\label{binary_2}\\
    \hat{B}_{1} &=&  \left(\sum_{i=Y+1}^M \hat{E}_i^\dag\hat{E}_i\right)^{1/2}\hat{R}^{-1/2} + \hat{Q},\label{binary_3}
\end{eqnarray}
with $\hat{R}=\sum_{i=1}^M \hat{E}_i^\dag\hat{E}_i$. Similar to Eq.~\eqref{one_out}, $\hat{R}^{-1/2}$ denotes the Moore-Penrose pseudoinverse and $\hat{Q}$ is the identity projector in the null space of $\hat{R}$ \cite{binary2008}. 
Since our POVMs for CV-to-DV entanglement conversion are diagonal in Fock basis, the binary measurement operators will also be Fock diagonal. If $B_0$ is obtained, the outcome belongs to one in the first group, otherwise the outcome belongs to the second. After the binary measurement, the number of outcomes need to be distinguished are reduced.  The process continues until a specific outcome is identified.

In Fig.~\ref{tree2} (b), we demonstrate the group division method with the entanglement transformation of a TMSV with $\lambda =0.8$ to a maximally entangled qubit pair. In descending order of probabilities, the corresponding POVMs of the first few outcomes are $\hat{E}_1 = \hat{M}_{02}, \hat{E}_2 = \hat{M}_{03}, \hat{E}_3 = \hat{M}_{01},\hat{E}_4 = \hat{M}_{15},\hat{E}_5 = \hat{M}_{14},\hat{E}_6 = \hat{M}_{04}$ and their probabilities are $P(E_1)=P_{02}=0.295,P(E_2)=P_{03}=0.188,P(E_3)=P_{01}=0.181,P(E_4)=P_{15}=0.0773,P(E_5)=P_{14}=0.0652,P(E_6)=P_{04}=0.0555$ respectively. In the first round of division, the first two events $\{E_1,E_2\}$ are assigned to the binary outcome $B_0$, such that their probabilities sum, labeled beside the branches in Fig.~\ref{tree2} (b), satisfies Eq.~\eqref{eq_prob_div}. The remaining outcomes, $\{E_3,E_4...\}$, are assigned to binary outcome $B_1$. If the first binary measurement gives $B_0$, the next round of binary measurement will distinguish the outcome $E_1$ from $E_2$. Otherwise, we divide the remaining outcomes into two groups, which in this case are $\{E_3, E_4\}$ and $\{E_5, E_6, ...\}$. The process continues until we obtain a binary measurement outcome that contains only one outcome $E_m$.

\subsection{Efficiency of Binary Search}

We assess the efficiency of the binary search schemes by considering the average number of measurement rounds needed to obtain one ebit of maximal DV entanglement from a TMSV state,
\begin{eqnarray}\label{eff}
    \eta \equiv \braket{R}/S.
\end{eqnarray} 
Here, the denominator $S$ is the average amount of maximal DV entanglement, in the unit of ebit, obtained per entanglement transformation. For the maximal qubit scheme in Sec.~\ref{sec_TMSV_To_Bell}, $S$ will be the success probability, $S=P_{\text{max}}$. For the random qudit scheme in Sec.~\ref{scheme2}, $S$ will be the average obtained entanglement in Eq.~\eqref{s_avg}, $S=S_{avg}$. The numerator is the expected number of measurement rounds for the entanglement transformation to complete, $\braket{R} = \sum_{n=1}^\infty P(E_n) R(E_n)$, where $R(E_m)$ denotes the number of measurement rounds required to distinguish the outcome $E_m$.

To appraise the performance of our binary search method, we consider a lower bound of binary measurement rounds that is necessary but might not always be attainable. Our idea is to treat the measurement sequence as a process to reveal the identity of a random variable that has the identities $E_m$ with probability $P(E_m)$. The minimum information required to describe this variable is thus given by the Shannon entropy $\text{H}=\sum_{n=1}^\infty P(E_n)\left[-\log_2P(E_n)\right]$ \cite{QCQI_textbbok}. Because each round of binary measurement can at most reveal $1$ bit of information, we can draw a lower bound of measurement round as $\braket{R} \geq \text{H}$. Since our near-even scheme attempts to increase the information revealed in each round by making the binary outcomes more evenly probable, we expect the performance of our scheme will be closer to the bound than the simple one-outcome-per-round scheme.

\begin{figure}
    \centering
    \includegraphics[scale=0.25]{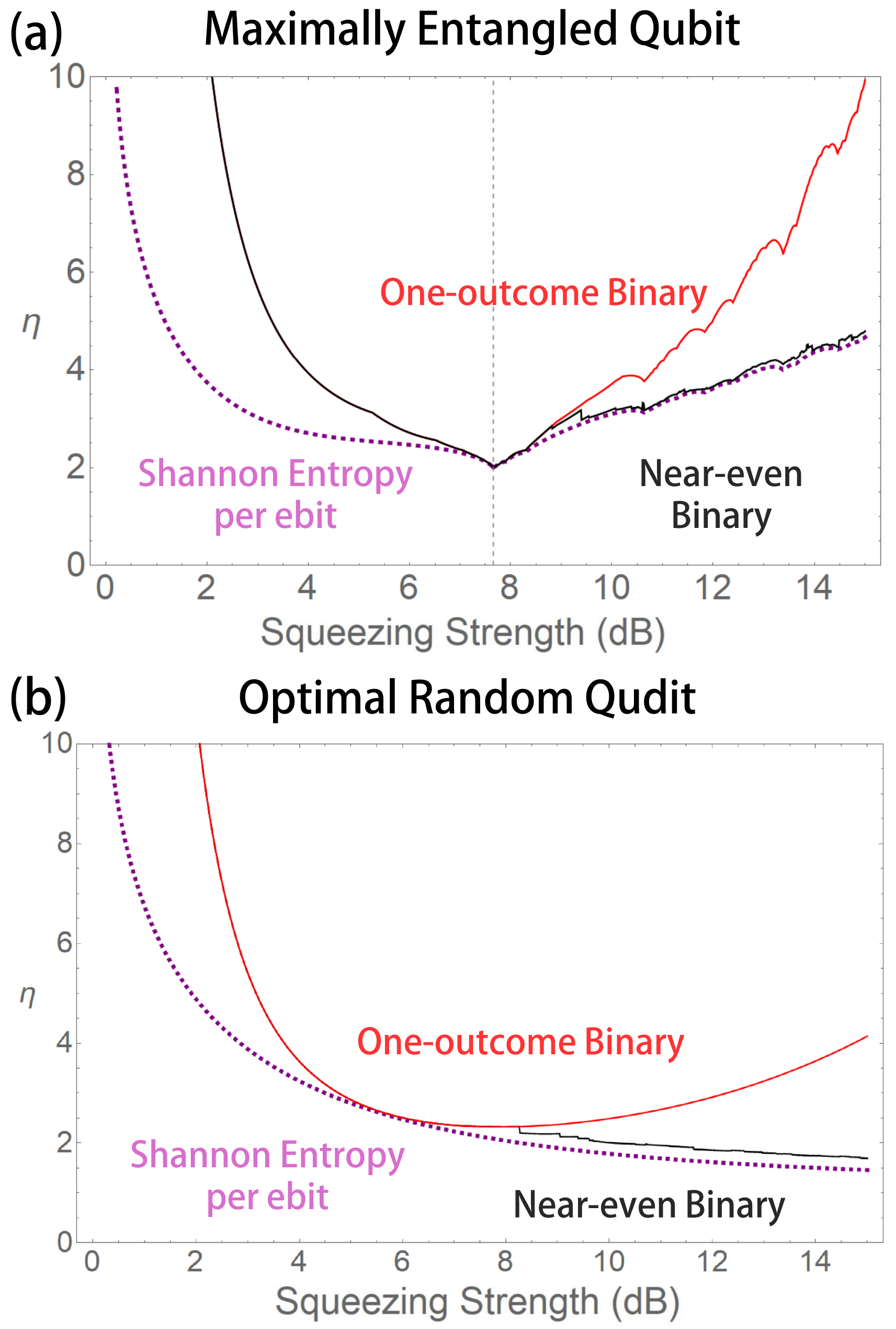}
    \caption{Average number of binary measurement rounds required to obtain 1 ebit of maximal DV entanglement at different squeezing strengths. Black and red curves respectively refer to the efficiencies of our near-even and the simple one-outcome-per-round schemes. The lower bound imposed by Shannon entropy is shown by a purple dashed for comparison. (a) Schemes for obtaining a maximally entangled qubit pair. (b) Schemes for obtaining randomly a maximally entangled qudit pair.}
    \label{compare}
\end{figure}

Fig.~\ref{compare}(a) compares the efficiencies for the maximal qubit scenario in Sec.~\ref{sec_TMSV_To_Bell}. In the probabilistic regime, $\lambda<1/\sqrt{2}$, both schemes coincide because the near-even scheme selects $Y=1$ for every group division, reducing to the one-outcome-per-round scheme. When only success outcomes remain, the largest probability $P_{0n}$ exceeds the sum of all other remaining probabilities, so it is isolated in the next binary measurement. When the failure outcome is also included, the two largest remaining probabilities still sum to more than half of the total, ensuring $Y=1$ according to Eq.~\eqref{eq_prob_div}. In the deterministic regime, the near-even scheme significantly outperforms the one-outcome-per-round scheme and approaches the lower bound over a wide range of squeezing strengths. Remarkably, at the threshold of $7.66$ dB, both schemes attain the lower bound. In this situation, a maximally entangled qubit pair can be obtained with two rounds of measurement on average, even though the original TMSV contains infinite non-zero Schmidt coefficients.

For the random-qudit scenario in Sec.~\ref{scheme2}, the efficiencies are shown in Fig.~\ref{compare} (b). The two schemes again coincide for squeezing strengths below approximately $7.66$ dB, since the grouping rule in Eq.~\eqref{eq_prob_div} selects $Y=1$ at every round. It reproduces the one-outcome-per-round scheme. At around $5.5$ dB, both schemes approach the lower bound imposed by Shannon entropy. Above $7.66$ dB, the near-even binary search can group multiple outcomes, reducing the average number of measurement rounds.

\section{Physical Implementation of POVMs}\label{phyiscal}

To realize our schemes, Alice is required to implement the Fock-diagonal binary measurement operators
\begin{eqnarray}\label{binary}
    \hat{B}_{0} &=& \sum_{n=0}^\infty C_{n}\ket{n}_A\bra{n}, \nonumber\\
    \hat{B}_{1} &=& \sum_{n=0}^\infty \sqrt{1-C_n^2}\ket{n}_A\bra{n},
\end{eqnarray}
where the coefficients $C_{n}$ are determined by Eqs.~\eqref{binary_2} and \eqref{binary_3}. Here, we illustrate a possible physical implementation of the measurement operators \eqref{binary} by coupling the bosonic mode to an ancillary qubit and then measuring the qubit.

We initialize the qubit in $\ket{g}$ and then apply the boson-number-dependent rotation
\begin{eqnarray}
    \ket{n}_A\ket{g} \to C_{n}\ket{n}_A\ket{g}+ \sqrt{1-C_{n}^2}\ket{n}_A\ket{e},
\end{eqnarray}
where $\ket{g}$ and $\ket{e}$ are the qubit computational basis states. This transformation can be realized using a photon-number-selective qubit rotation (SQR) gate \cite{PhysRevApplied.15.044026}, $\hat{U}_{\text{SQR}}\equiv \sum_n \exp(-i\phi_n\hat{\sigma}_y)\ket{n}\bra{n}$, where the rotation angles are $\phi_n =\arccos(C_n)$ and $\hat{\sigma}_y \equiv i \ket{e}\bra{g}-i \ket{g}\bra{e}$. Such SQR gates have been implemented extensively in circuit QED platforms using multi-tone drives, with each tone addresses a qubit frequency that is shifted by a specific boson number state \cite{PRXQuantum.3.030301, PhysRevLett.133.260802}. It can also be engineered by using quantum signal processing techniques on hybrid qubit-oscillator platforms that exhibit dispersive or Jaynes–Cummings couplings \cite{qsp_gate_2025}.

Following the above transformation, the auxiliary qubit is then projectively measured in the computational basis. The measurement outcome $\ket{g}$ heralds the implementation of $\hat{B}_{0}$, while the outcome $\ket{e}$ corresponds to $\hat{B}_{1}$.

\section{Conclusion}\label{conclusion}

We have developed optimal schemes for converting realistic CV bipartite entanglement into maximal DV entanglement. Two optimal scenarios were analyzed: (i) converting a TMSV state into a maximally entangled qubit pair at the theoretically maximal success rate, and (ii) converting into a maximally entangled qudit pair with random dimension that maximizes the average entanglement. In the first scenario, we identified the relation between the maximum success rate and the squeezing strength of TMSV, and found that none of the existing CV-to-DV entanglement conversion schemes is optimal. For the second scenario, we identified the optimal probabilities of obtaining each maximally entangled qudit pair and analytically found the entanglement sacrificed during the conversion.

For both scenarios, we constructed the measurement operators that realize the optimal schemes. Among known methods of entanglement conversion, we found that the method of areas can accomplish the task with the smallest number of POVMs. We further developed a binary tree scheme to reduce the number of rounds of binary measurement to implement the POVMs. Although the number of required POVMs is infinite, our schemes can accomplish the conversion in a finite number of measurement rounds on average. These techniques make CV-to-DV entanglement conversion experimentally feasible and resource-efficient.

Finally, we illustrated how the measurement operators can be implemented in hybrid qubit-oscillator platforms. Our work establishes a practical pathway to optimally generate broadly applicable DV entanglement from the efficiently produced CV entanglement. It opens new opportunities for hybrid quantum technologies that combine the advantages of both continuous- and discrete-variable quantum information, and facilitates the implementation of quantum technologies on physically advantageous bosonic platforms.

\begin{acknowledgments}
This work is supported by the Natural Sciences and Engineering Research Council of Canada Discovery Grant (NSERC RGPIN-2021-02637), Alliance International Catalyst Quantum Grant (ALLRP 578638-22), and Canada Research Chairs (CRC-2020-00134). R. Tullu acknowledges support from an NSERC USRA. The authors acknowledge the use of ChatGPT (OpenAI) for assistance with manuscript preparation, including language editing and checks of mathematical and logical consistency. The authors remain fully responsible for the accuracy and scientific content of this work.
\end{acknowledgments}

\begin{appendix}

\section{Conversion Rate of Different Schemes}\label{app_conversion_rate}

Fig.~\ref{optimal_rate} (b) and (c) compare the conversion rates, i.e. the success probability of obtaining one maximally entangled qubit pair per TMSV resource state, for the schemes proposed in the literature and the current work. In this appendix, we present the explicit expression of the conversion rates of these schemes. 

\subsection{SPDC Scheme}

In optical systems, SPDC is a common entanglement resource. As discussed in Sec.~\ref{sec_optimal_rate}, a maximally entangled photonic qubit pair can be probabilistically obtained from two pairs of SPDC photons \cite{PhysRevLett.98.190503, Krovi2016}. The success probability is given by, the squared overlap between the total state in Eq.~\eqref{spdc} and the target state, $(\ket{01}_s\ket{01}_i+\ket{10}_s\ket{10}_i)/\sqrt{2}$, 
\begin{eqnarray}
    P_{\text{SPDC}} = 2\lambda^2(1-\lambda^2)^2.
\end{eqnarray}
Because this protocol consumes two TMSV states per successful conversion attempt, the conversion rate is  
\begin{eqnarray}
    \text{Conversion Rate} = \frac{P_{\text{SPDC}}}{2}=\lambda^2(1-\lambda^2)^2.
\end{eqnarray}
In the weak-squeezing regime, $\lambda\ll 1$, it is approximately half of the optimal rate, $P_{\text{max}}=2\lambda^2$.

\subsection{NLA Scheme}

Another photonic approach is NLA \cite{PhysRevA.102.063715}. In this scheme, Alice's mode interferes alternatively with two ancillary modes, one prepared in a single-photon state and the other in vacuum. Conditioned on heralding one ancillary mode in the single-photon state and the other in vacuum, Alice's mode is truncated to the single-photon subspace while modifying the vacuum coefficient. Thus, the input TMSV state is projected onto $(\lambda\ket{0}_A\bra{0}+\ket{1}_A\bra{1})\ket{\text{TMSV}}$. The success probability is equal to the square of the projection between the final state, $\ket{\text{Bell}_{01}}=(\ket{0}_A\ket{0}_B+\ket{1}_A\ket{1}_B)/\sqrt{2}$, and the projected state, 
\begin{eqnarray}
    P_{\text{NLA}} &=& |\braket{\text{Bell}_{01}|(\lambda\ket{0}_A\bra{0}+\ket{1}_A\bra{1})|\text{TMSV}}|^2\nonumber\\
    &=&2(1-\lambda^2)\lambda^2,
\end{eqnarray}
which is nearly optimal in the weak-squeezing regime. Since this scheme, as well as the remaining schemes discussed below, requires only one TMSV state per conversion attempt, its conversion rate is simply equal to the success probability.

\subsection{Scheme Proposed in Ref.~\cite{PhysRevA.72.034101}}

A related interference-and-detection scheme proposed in Ref.~\cite{PhysRevA.72.034101} can also project a TMSV onto $\ket{\text{Bell}_{01}}$ without requiring a single-photon ancilla. However, its success probability is lower because it must eliminate the dominant vacuum component of the TMSV. In this scheme, each bosonic mode interferes with a vacuum ancillary mode through an unbalanced beam-splitter. Physically, interference with vacuum and detecting a single photon effectively corresponds to removing one photon from the state. By adjusting the reflection/transmission of the beam-splitter, one obtains $\ket{\text{Bell}_{01}}$ in the weak-squeezing regime, conditioned on both ancillary modes being projected onto the non-vacuum subspace. The corresponding success probability is
\begin{eqnarray}
    P_{\text{w/o}} = \frac{(1-\lambda^2)^{5/2}(2+4\lambda^2-2\sqrt{1+4\lambda^2})}{16}.
\end{eqnarray}

\subsection{Superposition of Photon Addition and Subtraction Scheme}

One can also apply the superposition of photon addition and subtraction  to each mode of a TMSV state, $(T\hat{a}+R\hat{a}^\dag)(T\hat{b}+R\hat{b}^\dag)$, to prepare an approximate maximally entangled photonic qubit pair \cite{PhysRevA.84.012302,PhysRevA.86.012328,PhysRevLett.101.260401, PhysRevA.82.053812}. This operation gives 
\begin{eqnarray}\label{subtraction_state}
    &&\sqrt{\mathcal{A}}  (T\hat{a}+R\hat{a}^\dag)(T\hat{b}+R\hat{b}^\dag)\ket{\text{TMSV}}\nonumber\\
    =&& \sqrt{\mathcal{A}(1-\lambda^2)} \Big( R^2\ket{1}_A\ket{1}_B  \nonumber\\
    &&+\lambda (T^2+2R^2)\ket{1_\perp}_A\ket{1_\perp}_B+ O(\lambda^2) \Big),
\end{eqnarray} 
where
\begin{eqnarray}
    \ket{1_\perp} = \frac{T\ket{0}+\sqrt{2}R\ket{2}}{\sqrt{T^2+2R^2}},
\end{eqnarray}
which is orthogonal to $\ket{1}$ and the normalization factor is
\begin{eqnarray}
    \mathcal{A} = \frac{(1-\lambda^2)^2}{\lambda^2(1+4R^2T^2)+(R^2+\lambda^2T^2)^2}.
\end{eqnarray}
Here $\hat{a}$ and $\hat{b}$ are the annihilation operators for Alice's and Bob's modes, respectively, and $R, T \in \mathbb{R}$ satisfy $R^2+T^2=1$.

Physically, this operation can be implemented probabilistically by coherently superposing two heralded processes. Photon subtraction is obtained by tapping off a small fraction of the input field with a highly transmissive beam splitter and conditioning on a photon detection in the reflected mode; in the weak-reflection limit, the transmitted state is acted on by $\hat{a}$. Photon addition is obtained from weak parametric down-conversion, where detection of an idler photon heralds the action of $\hat{a}^\dagger$ on the signal mode. If these two heralding processes are made indistinguishable in an interferometric arrangement, the conditional operation becomes the coherent superposition $T\hat{a}+R\hat{a}^\dagger$, with $R$ and $T$ controlled by the beam-splitter ratios and phases. Choosing
\begin{eqnarray}\label{R_lambda}
    R=\sqrt{\frac{\lambda}{1-\lambda}}
\end{eqnarray}
makes the coefficients of the first two terms in Eq.~\eqref{subtraction_state} equal. Thus, in the weak-squeezing regime, $\lambda \ll 1$, the state in \eqref{subtraction_state} contains a maximally entangled qubit pair in the subspace spanned by $\{\ket{1},\ket{1_\perp}\}$ on each mode.

By filtering the states outside the $\{\ket{1}, \ket{1_\perp}\}$ in both modes, for example by the POVM $\{\ket{1}\bra{1} + \ket{1_\perp}\bra{1_\perp}, \text{else}\}$, the TMSV state will be projected to the unnormalized successful state
\begin{eqnarray}
    \ket{\psi_{\mathrm{success}}} = \lambda\sqrt{1-\lambda^2}
\left(\ket{1}_A\ket{1}_B + \ket{1_\perp}_A\ket{1_\perp}_B\right).\nonumber\\
\end{eqnarray}
The success probability of this filtering is given by the squared norm of this state,
\begin{eqnarray}
    P_{s/a}
=\braket{\psi_{\mathrm{success}}|\psi_{\mathrm{success}}}
=2\lambda^2(1-\lambda^2).
\end{eqnarray}
Thus, the successful probability of this protocol also approaches the optimal conversion probability $P_{\max}=2\lambda^2$ in the weak-squeezing regime.

\subsection{Entanglement Download Scheme}

For hybrid qubit-oscillator systems, we explore the entanglement-downloading scheme proposed in Ref.~\cite{han2025download}. This scheme converts the correlations embedded in an $N_m$-mode CV cluster state into entanglement among $N_m$ qubits, thereby producing a qubit cluster state with the same underlying graph. In the protocol, each oscillator is coupled to a qubit through a conditional displacement gate, and then the oscillator is measured by homodyne detection. The protocol can be understood as teleporting the qubit correlations hidden in the displaced-GKP decomposition \cite{PhysRevA.64.012310,PhysRevA.73.012325} of the CV cluster state into physical qubits.
In the two-mode case, $N_m=2$, the CV cluster state is equivalent to a TMSV state up to certain local Gaussian unitaries \cite{giedke2003entanglementtransformationspuregaussian}, while the downloaded two-qubit cluster state is equal to a Bell state up to local Clifford unitaries.

For an ideal CV cluster state, this procedure deterministically transfers the cluster-state entanglement to the qubits. However, an ideal CV cluster state is unphysical because it requires infinite squeezing, and hence infinite energy, on each mode. For a realistic finitely-squeezed CV cluster state, the downloaded state is not an ideal cluster state. The imperfection can be probabilistically corrected by a local weak measurement on each qubit. Upon success, the corresponding qubit is restored to the correct cluster state; otherwise, the qubit is disentangled with other qubits. Following Ref.~\cite{han2025download}, the average success probability for each qubit is 
\begin{eqnarray}
    p_{\text{suc}}=1-\text{erf}\left(R_0\sqrt{\frac{\pi}{2}}\right),
\end{eqnarray}
where $R_0 \geq 0$ quantifies the finitely-squeezing imperfection of the CV cluster state and vanishes in the infinite-energy limit.

In the two-mode scenario considered here, $R_0$ is related to the two-mode squeezing parameter $\lambda$ by
\begin{eqnarray}
    R_0= \sqrt{\frac{1-\lambda^2}{2\lambda^2}}.
\end{eqnarray}
Since the downloaded two-qubit cluster state is regarded as successfully prepared only when both local correction operations succeed, the total success probability of the CV-to-DV conversion is 
\begin{eqnarray}
    P_{\text{download}} &=&  p_{\text{suc}}^2\nonumber\\
    &=&\left(1-\text{erf}\left(\sqrt{\frac{1-\lambda^2}{2\lambda^2}}\sqrt{\frac{\pi}{2}}\right)\right)^2.\nonumber\\
\end{eqnarray}

\section{Scaling of the number of POVMs in Nielsen's method}\label{app_NMethod}

In Fig.~\ref{number_of_operators}, we compare the number of POVMs, which is also equal to the number of measurement outcomes, required by different entanglement-transformation schemes. In particular, Nielsen proposed a scheme to realize deterministic pure-state entanglement transformation through a sequence of binary measurements \cite{Nielsen1999} and it was later extended by Vidal for the probabilistic transformation \cite{Vidal1999}. In this appendix, we explain how these schemes apply to CV-to-DV entanglement conversion, and why their implementation requires an exponentially large number of POVMs.

Deterministic conversion is possible when the squared-Schmidt coefficients of the target state majorize those of the initial, as expressed in Eq.~\eqref{cumulative probabilities}. If $\vec{\alpha}$ and $\vec{\beta}$ denote the squared-Schmidt coefficient vectors of the initial and target states, respectively, the majorization condition is equivalent to the existence of a doubly stochastic matrix $T$ that relates the two vectors as $\vec{\alpha}=T \vec{\beta}$. Nielsen showed that $T$ can be decomposed into a product of $r$ $T$-transforms $D_m$,
\begin{equation}
    T   = D_rD_{r-1}...D_2 D_1.
\end{equation}
Each $T$-transform mixes only two components of the vector and has the form
\begin{equation}
    D_m=t_m I+(1-t_m)\Upsilon_m,
    \qquad
    0\leq t_m\leq 1,
\end{equation}
where $\Upsilon_m$ is a permutation matrix that exchanges a particular pair of components.

The number of required $T$-transforms, $r$, follows from the standard constructive proof of majorization. Each $T$-transform is added to fix one additional component of the Schmidt vector while leaving the components that have already been fixed unchanged. Hence, for an $N$-dimensional initial vector, at most $r=N-1$ $T$-transforms are required. If some Schmidt coefficients of the initial and target states coincide, fewer transformations may be sufficient, so that $r\leq N-1$.

Each $T$-transform is implemented by one binary measurement on Alice's side. 
For example, consider the intermediate step $\vec{a}=D_m\vec{b}$, where $\vec{b}\equiv D_{m-1}...D_1\vec{\beta}$. Suppose that $D_m$ related mixes the $l$-th and $l'$-th Schmidt components, the corresponding binary measurement consists of the measurement operators
\begin{eqnarray}
    \hat{\mathbb{B}}_{0,m} &=& \sqrt{t_m}\Big(\sum_{n\neq l,l'}\ket{w_n}_A\bra{w_n} \nonumber\\
    &+& \sqrt{\frac{b_l}{a_l}}\ket{w_l}_A\bra{w_l} + \sqrt{\frac{b_{l'}}{a_{l'}}}\ket{w_{l'}}_A\bra{w_{l'}}\Big), \nonumber\\
   \hat{\mathbb{B}}_{1,m} &=& \sqrt{1-t_m}\Big( \sum_{n\neq l,l'}\ket{w_n}_A\bra{w_n},\nonumber\\
    &+& \sqrt{\frac{b_{l}}{a_{l'}}}\ket{w_{l}}_A\bra{w_{l'}} + \sqrt{\frac{b_{l'}}{a_{l}}}\ket{w_{l'}}_A\bra{w_{l}}\Big), \nonumber\\
\end{eqnarray}
where $\ket{w_{l}}$ denote the $l$-th Schmidt basis state. Thus, each $T$-transform gives rise to two possible measurement outcomes. After $r$ rounds of binary measurements, the total number of measurement outcomes is $2^r$.

We now apply this counting to CV-to-DV conversion. The initial state is a truncated TMSV state, up to $N-1$ excitations in each mode, while the target state is a maximally entangled qubit pair. Since the Schmidt basis is the Fock state basis, $\ket{w_n}_A=\ket{n}_A$, the converted qubit state is the equal superposition of Fock states \eqref{eq_bell}. 
In general, for $\lambda>1/\sqrt{2}$, the Nielsen's scheme requires $r=N-1$ binary measurement rounds, giving $2^{N-1}$ measurement outcomes. At the threshold point $\lambda=1/\sqrt{2}$, the largest Schmidt coefficient of the initial state coincides with that of the target state, so one fewer $T$-transform is required to match the Schmidt coefficients. The protocol then requires $r=N-2$ rounds of binary measurement and yields $2^{N-2}$ measurement outcomes.

For the probabilistic case, the strategy is to first project the initial state onto an intermediate state that satisfies the majorization condition \cite{Vidal1999}. Here we choose the intermediate state to be 
\begin{eqnarray}
    \frac{1}{\sqrt{2}}\ket{0}_A\ket{0}_B+\sqrt{\frac{1-\lambda^2}{2}}\sum_{n=1}^\infty \lambda^{n-1}\ket{n}_A\ket{n}_B.
\end{eqnarray}
This heralded projection can be implemented by the binary measurement
\begin{eqnarray}
     \hat{\mathbb{B}}_{0} &=& \frac{\lambda}{\sqrt{1-\lambda^2}}\ket{0}_A\bra{0}+\sum_{n=1}^\infty \ket{n}_A\bra{n}, \nonumber\\
    \hat{\mathbb{B}}_1 &=& \sqrt{\frac{1-2\lambda^2}{1-\lambda^2}}\ket{0}\bra{0}.
\end{eqnarray}
When the outcome associated with $\hat{\mathbb{B}}_{0}$ is obtained, the projection succeeds and the resulting state can subsequently be converted deterministically into a maximally entangled qubit pair using the deterministic scheme described above. Since the largest squared Schmidt coefficient of the intermediate state is $1/2$ and coincides with that of the target state, the deterministic conversion requires only $r=N-2$ rounds of binary measurement. Considering the intermediate binary measurement has one successful outcome that continues into the deterministic protocol and one failure outcome that terminates, the total number of the possible outcomes is $2^{N-2}+1$.

Thus, both deterministic and probabilistic Nielsen-type CV-to-DV conversion protocols require $O(2^N)$ POVMs.

\section{Construction of the transformed Chart}\label{app_method_area}

In this appendix, we describe the area construction used in Fig.~\ref{method_area}.  The construction is based on Hardy's method of areas \cite{Hardy1999}, but we use a simplified variant adapted to the CV-qubit conversion considered in this work.

In Hardy's construction, each Schmidt coefficient of the initial state is represented as a vertical column, and each column is assigned a distinct colour. The colour keeps track of the corresponding Schmidt label on Bob's system. The columns are then divided into patches, which can be rearranged through relabeling the basis of Alice's system. The goal of the rearrangement is to form a transformed chart consisting of a success region, corresponding to the desired target states, and possibly a failure region, corresponding to unsuccessful outcomes. Examples for the above- and below-threshold CV-qubit entanglement conversion are shown in Figs.~\ref{method_area}(b) and (c), respectively. The physical constraint of entanglement conversion can be represented by a colouring constraint: no horizontal level of the transformed chart contains two patches of the same colour. This condition ensures that every component in the conditional output state is associated with distinct Schmidt labels on Bob's side.

Hardy's original construction gives a general ordered cut-and-paste procedure that preserves the colouring constraint. The procedure begins by identifying the regions of area to be removed from the initial chart and the regions to be filled in the target chart. Starting from the right, a patch is taken from the rightmost region to be removed and placed as high as possible in the rightmost region to be filled. If the patch is larger than the unfilled space in the target column, its excess portion is exchanged with an equal amount of area already occupying that column. The exchanged area is moved to the original position of the inserted patch and subsequently transferred to the next region to be filled. This procedure is repeated until the target chart is obtained. The construction used below follows the same logic, but incorporates simplifications that are tailored for the CV-qubit entanglement conversion.

Since our target state is a maximally entangled qubit pair, the success region of the transformed chart is two-column.  At each horizontal level, there are at most two patches, and the colouring constraint reduces to the requirement that these two patches have different colours. This allows us to adopt a simplified construction.

\begin{figure}
    \centering
    \includegraphics[scale=0.23]{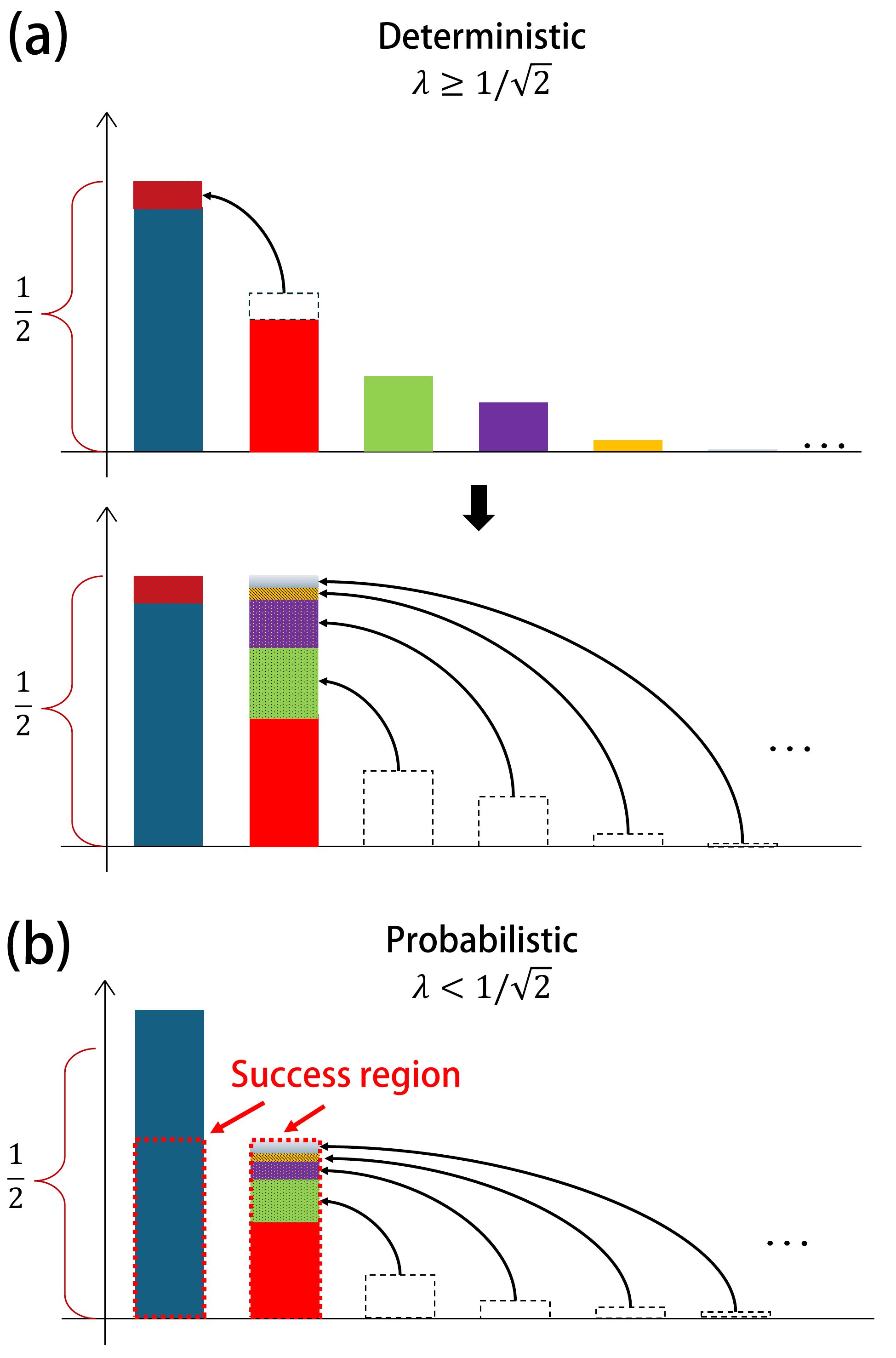}
    \caption{Illustrative examples of the simplified method of areas procedure. The arrows indicate the transfer of coloured patches, while the dashed outlines indicate their original positions. (a) In the deterministic regime, patches are successively transferred from the second and subsequent columns to complete the first target column to height $1/2$. If the last patch exceeds the target height, it is divided into two parts: one fills the first column on top and the excess is retained at the bottom of the second target column. The remaining patches are then stacked to complete the second target column to the same height of $1/2$. (b) In the probabilistic regime, the second and subsequent columns are stacked into the second target column. The red dotted outlines mark the corresponding success region, which consists of two equal-height columns and comprises the entire second column and the portion of the first column up to the same height.}
    \label{area_app}
\end{figure}

We begin with the initial area chart shown in Fig.~\ref{method_area}(a). In the deterministic regime, the goal is to form two columns of height $1/2$. In the first step, patches from the second column and onward are successively moved to the first target column, stacking them above the existing patches until its height reaches $1/2$. As illustrated in the upper panel of Fig.~\ref{area_app}(a), if the last patch exceeds the remaining space, it is divided into two parts: one is placed on the top of the first target column, completes it, while the other is placed at the bottom of the second target column. In the second step, all remaining patches are then stacked onto the second target column, as illustrated in the lower panel of Fig.~\ref{area_app}(a). Since the total area is conserved, the second target column also reaches height $1/2$, yielding the transformed chart shown in Fig.~\ref{method_area}(b).

The colouring constraint is preserved by this construction. Each original colour appears in only one target column, except possibly the colour of the last patch used to complete the first target column. If this patch is divided, its two parts occupy the top of the first target column and the bottom of the second target column, respectively. Since every original column has height no greater than $1/2$ in the deterministic regime, these two parts cannot overlap at the same horizontal level. Consequently, each horizontal slice of the completed two-column block contains two different colours and corresponds to a valid outcome of maximally entangled qubits. The different horizontal intervals in Fig.~\ref{method_area}(b) represent different measurement outcomes, with probabilities given by the areas of the corresponding horizontal strips.

In the probabilistic regime, the first original column exceeds height $1/2$, so moving and stacking patches from the right cannot construct two columns with equal height of 1/2, thus deterministic creation of maximally a entangled qubit pair is impossible. Instead, we construct a smaller success region with two equal-height columns, as shown in Fig.~\ref{method_area}(c). The second target column is formed by successively stacking all patches from the second and subsequent original columns, as illustrated in Fig.~\ref{area_app}(b). A portion of the first original column, with the same height as the completed second target column, forms the other side of the success region. The remaining upper portion of the first column corresponds to the failure outcome.

The colouring constraint is also preserved in this case. The first target column contains only the colour of the first original column, whereas the second target column consists entirely of patches from the remaining original columns. Hence, the two patches appearing at each horizontal level of the success region always have different colours, ensuring that each horizontal slice corresponds to a valid outcome of maximally entangled qubits.

We remark that this simplified construction is applicable to the present CV-qubit entanglement conversion because it preserves the colouring constraint throughout the two-column success region. For more general target states involving higher Schmidt ranks or different output distributions, the same stacking procedure may not preserve the colouring constraint, and Hardy's more general construction may be required.

\section{Constant Entanglement Loss from the TMSV state to the Optimal Random Qudits in the Large-Squeezing Limit}\label{app_constant}

In Fig.~\ref{average}, we compare the entanglement of the original TMSV state and the converted maximally entangled DV states. Interestingly, as the squeezing increases, the difference between the entanglement entropy of the initial TMSV and the average DV entanglement produced by the optimal random qudit scheme approaches a constant. In this appendix, we derive this asymptotic constant.

The average DV entanglement of the optimal random qudit scheme is given in Eq.~\eqref{s_avg}. To analyze its behavior in the strong-squeezing limit, $\lambda\to1$, we write $\lambda^2\equiv e^{-\mu}$, so that the limit becomes $\mu\to0$. We then expand Eq.~\eqref{s_avg} in powers of $\mu$.

The relevant asymptotic expansion follows from the Mellin transform \cite{book_mathematical_method} 
\begin{eqnarray}\label{Li_complicated}
    \text{Li}_s(e^{-\mu}) &\equiv&  \sum_{n=1}^\infty \frac{e^{-n\mu}}{n^s}\nonumber\\
   & =& \sum_{n=1}^\infty \frac{1}{n^s}\frac{1}{2\pi i} \int^{c+i\infty}_{c-i\infty} \Gamma(t)(n\mu)^{-t}dt \nonumber\\
   &=&  \frac{1}{2\pi i} \int^{c+i\infty}_{c-i\infty} \Gamma(t)\mu^{-t} \zeta(t+s)dt, 
   \end{eqnarray}
where we used the inverse Mellin transform of the exponential function, i.e. $e^{-x} = \int^{c+i\infty}_{c-i\infty} \Gamma(t)x^{-t}dt/2\pi i$ with $c>\text{max}(1-s,0)$, on the second line. Here $\Gamma(t)$ is the Gamma function and $\zeta(t)=\sum_{n=1}^\infty n^{-t}$ is the Riemann zeta function \cite{book_mathematical_method}.

The integrand has a simple pole at $t=1-s$ from $\zeta(t+s)$ and simple poles at $t=0,-1,-2,...$ from $\Gamma(t)$. Evaluating the residues gives the expansion
\begin{eqnarray}
    \text{Li}_s(e^{-\mu}) = \Gamma(1-s)\mu^{s-1} + \sum_{k=0}^\infty \zeta(s-k)\frac{(-\mu)^k}{k!},\nonumber\\
\end{eqnarray}
where we used the identity $\text{Res}_{t=-k} \Gamma(t)=(-1)^k/k!$ for integers $k\geq0$.

Differentiating with respect to $s$ gives
\begin{eqnarray}
    \frac{\partial \text{Li}_s(e^{-\mu})}{\partial s}&=& \Gamma(1-s)\mu^{s-1}(\ln\mu-\psi(1-s)) \nonumber\\
    &+& \sum_{k=0}^\infty \frac{\partial}{\partial s}\zeta(s-k)\frac{(-\mu)^k}{k!},
\end{eqnarray}
where $\psi(s)$ is the digamma function \cite{polylogarithm_book}. At $s=-1$, the leading term in the strong-squeezing limit is
\begin{eqnarray}
    \frac{\partial \text{Li}_s(e^{-\mu})}{\partial s}|_{s=-1} = \Gamma(2)\mu^{-2}(\ln\mu-\psi(2))+O(1),\nonumber\\
\end{eqnarray}
where $\psi(2)=1-\gamma$, $\Gamma(2)=1$, $\gamma$ is the Euler-Mascheroni constant \cite{euler_book}.

In Eq.~\eqref{s_avg}, this derivative is multiplied by 
\begin{eqnarray}
    \frac{(1-\lambda^2)^2}{\lambda^2}=\mu^2 +O(\mu^4).
\end{eqnarray}
Therefore, the average entanglement has the asymptotic form
\begin{eqnarray}
    S_{\text{avg}} =\frac{-\ln\mu+1-\gamma}{\ln2} + O(\mu^2|\ln\mu|).
\end{eqnarray}

We can compare this with the entanglement entropy of the initial TMSV state. In terms of $\mu$, the TMSV entanglement is
\begin{eqnarray}
    S_{\text{TMSV}} 
    &=&-\log_2 (1-\lambda^2)-\frac{\lambda^2 \log_2\lambda^2}{1-\lambda^2}\nonumber\\
    &=&\frac{-\ln\mu+1}{\ln 2}+O(\mu^2|\ln\mu|).
\end{eqnarray}
Therefore, the asymptotic difference is
\begin{eqnarray}
    S_{\text{TMSV}}-S_{\text{avg}} = \frac{\gamma}{\ln2}+O(\mu^2|\ln\mu|),
\end{eqnarray}
and the constant difference is equal to $\gamma/\ln2\approx 0.832746$.

\end{appendix}

\bibliography{aps_reference}
\pagestyle{plain}

\end{document}